\documentclass[final,3p,times,authoryear]{elsarticle}

\usepackage{lineno,hyperref}
\modulolinenumbers[5]

\usepackage{parskip}
\usepackage[section]{placeins}
\usepackage[plain]{fancyref}
\usepackage{float}
\usepackage{array}
\usepackage{multirow}
\usepackage{adjustbox}
\usepackage{lscape}
\usepackage{bm}
\usepackage{adjustbox}

\usepackage[english]{babel}															% English language/hyphenation
\usepackage[protrusion=true,expansion=true]{microtype}				% Better typography
\usepackage{amsmath,amsfonts,amsthm,amssymb}									
\usepackage{graphicx}														
\usepackage{color,transparent}
\usepackage{multicol}													
\usepackage{epstopdf}																	
\usepackage{booktabs}																	 
\usepackage{textcomp}
\usepackage{textgreek}
\usepackage{gensymb}
\usepackage{rotating}
\usepackage{rotfloat}
\usepackage{subfig}
\usepackage[T1]{fontenc}
\usepackage{lmodern}

\graphicspath{ {images/} }

\biboptions{authoryear}

\begin{document}

\begin{frontmatter}
	
	
	
%%	\dochead{}
	
\title{Rheological behaviour and flow dynamics of Vitreous Humour substitutes used in eye surgery during saccadic eye movements}

\author[mymainaddress,mymainaddress2]{Andreia F. Silva \corref{mycorrespondingauthor}}
\cortext[mycorrespondingauthor]{Corresponding author}
\ead{andreia.silva@ed.ac.uk}
\author[mymainaddress3]{Francisco Pimenta}
\author[mymainaddress3]{Manuel A. Alves}
\author[mymainaddress]{M\'{o}nica S.N. Oliveira}
\address[mymainaddress]{James Weir Fluids Laboratory, Department of Mechanical and Aerospace Engineering, University of Strathclyde, Glasgow G1 1XJ, UK}
\address[mymainaddress2]{School of Physics and Astronomy, University of Edinburgh, King's Buildings, Mayfield Road, Edinburgh EH9 3JL, UK}
\address[mymainaddress3]{Departamento de Engenharia Qu\'{i}mica, CEFT, Faculdade de Engenharia, Universidade do Porto, 4200-465 Porto, Portugal}

\date{}
\journal{   }


\begin{abstract}
This work discusses the rheology of several vitreous humour (VH) substitutes used in eye surgery (perfluorocarbons and silicone oils) and their flow behaviour when subjected to saccadic eye movements. 
Shear rheology experiments revealed that all fluids tested exhibit a constant shear viscosity, while extensional rheological experiments showed that Siluron 2000 is the only fluid tested that exhibits a measurable elasticity. 
To characterise the dynamics during saccadic eye movements, numerical simulations of all the VH substitutes under study were performed with the open source software  OpenFOAM\textsuperscript{\textregistered} and compared with Vitreous Humour flow dynamics to assess their potential mechanical performance. Minor differences were found between the numerical results of a viscoelastic fluid reproducing the rheology of Siluron 2000 and a Newtonian model. Perfluorocarbon (PFLC) shows a distinct flow behaviour relative to Silicone Oils (SiO). None of the pharmacological fluids tested can adequately mimic the rheological and consequently the flow behaviour of VH gel phase \citep{Silva2020}. 
%The differences observed can explain, from a mechanical point of view, why those fluids cannot be used as a permanent substitute. 
\end{abstract}

\begin{keyword}
Vitreous humour substitutes, rheology, flow dynamics, saccadic movements
\end{keyword}

\end{frontmatter}

%\linenumbers

\section{Introduction}

Vitreous humour (VH) is a transparent gel-like avascular structure that fills the vitreous chamber (the space between the lens and the retina) in the eye, and is composed by water (approx. 99\%), salts and a network of collagen fibrils and hyaluronan (HA)  \citep{Chirila1998,Nickerson2005,Nickerson2008,Sharif-Kashani2011,Siggers2012}. Several functions are attributed to VH, including: playing an active role during growth to keep the retinal pigment epithelium in its proper position; providing a transparent medium for the passage of light to the retina; serving as a metabolic repository for hyalocytes and neighbouring tissues, and controlling the movement of solutes and solvents within the eye; also providing a system that absorbs mechanical stress and protects the surrounding tissues during eye movement and physical activity \citep{Lee1992}. 
Changes in the VH can have a serious impact in their functionality, eventually leading to the appearance of several diseases \citep{Baino2010,Lee1992}, such as: posterior vitreous detachment (PVD); retinal tears, retinal detachements and vitreomacular traction; diabetic retinopathy and vitreopathy; cataract formation; and age related macular degeneration.
\par
A common treatment for many of these diseases is the injection of a fluid in the vitreous cavity. Such substances, known as vitreous substitutes, can be used intra-operatively to push a detached retina to its normal position, to restore the volume of the vitreous cavity, and to help the surgeon in membrane dissection. Some of these agents can be used as post-operative adjuncts to provide internal tamponade after vitreous surgery \citep{Giordano1998}. Unfortunately, a substitute that can reproduce closely the properties of VH and remain effective for long periods or even permanently in the vitreous cavity has not yet been identified. The ideal vitreous substitute should mimic the native VH in both form and function while being easily manipulated during surgery.
\par
Currently, perfluorocarbons (PFLC) and silicone oil (SiO) based fluids are the only VH substitutes regulated and approved by the European Medicines Agency (EMA) - Europe, and the Food and Drug Association (FDA) - United States of America, for use in surgery in the posterior segment of the eye. 
\par
Perfluorocarbons (PFCLs) are clear, transparent and can dissolve gas, but are immiscible with aqueous solutions. PFCLs are mostly used in retinal detachments  for purely mechanical reasons, and are removed from the eye right after the retina is 
restored to its normal position. If left in the eye for long periods, PFCLs can induce morphological changes in the inferior retina caused by mechanical compression. 
Their high density and low viscosity, together with a low interfacial tension relative to the liquefied vitreous and the eye facilitate intraocular dispersion of PFCLs during head motion \citep{Giordano1998}. 
\par
The use of Silicone oils in this context started being studied in mid $20^{th}$ century. \cite{Stone1958} reported the first \textit{in vivo} study with SiO, in which liquid silicone was injected into the vitreous cavity of rabbits.
The results were promising and in 1962, the first clinical study with Humans was reported, with 33 patient eyes being injected with liquid silicone \citep{Cibis1962}. SiOs are arguably the most used VH substitutes showing suitable properties such as stability, transparency, and high interfacial surface energy with the eye tissues. The first SiOs used in eye surgery were composed by SiO molecules of one molecular weight (MW). Different viscosities were available, between 1 and 5 Pa s, corresponding to SiO  with MW between 25 and 50 kDa. 
SiO with lower viscosities have the advantages of requiring smaller needles and a shorter injection time during the surgical procedure \citep{Williams2010}. However, the emulsification rate of the SiO with lower viscosities is higher than that for the SiO with higher viscosity \citep{Light2006,Chan2011}. To try to overcome the emulsification problems, new substitutes composed of SiO with two different MWs were proposed. Adding a small amount of SiO with a high molecular weight (HMW) to a SiO with short chains was shown to increase the resistance of the fluid to emulsification \citep{Caramoy2010,Caramoy2011,Chan2011,Williams2010b,Caramoy2015}: the viscosity of the fluid increases when compared with the SiO with only short chains, but not enough to significantly increase the injection time during surgery \citep{Caramoy2011}. Note that ultimately, all the SiO fluids available for eye surgery need to be removed from the eye, independently of being composed by a single MW or by a mix of different MWs. If left for long periods, the presence of the fluids may lead to complications such as cataract, glaucoma, and keratopathy \citep{Barca2014,Baino2011,Giordano1998,Federman1988}.  

In previous works we characterised the rheological behaviour of Vitreous Humour \citep{Silva2017}, and simulated numerically the flow behaviour of VH liquid and gel phases during saccadic eye movements \citep{Silva2020}. Additionally, a review of the numerical work discussing the flow behaviour of VH fluid when subjected to saccadic movements can be found in \cite{Silva2020}. To the best of our knowledge there are no systematic studies on the flow behaviour of the different groups of vitreous substitutes, neither a  thorough comparison of their behaviour with the VH behaviour. Moreover, it is known that depending on the SiO MW, the fluid may or may not have elasticity. For the SiOs typically used in eye surgery, it is possible to find works that assume that they are viscoelastic \citep{Romano2009,Caramoy2010} and others that consider that they behave as Newtonian fluids \citep{Feng2013}. However, it is not possible to find in the literature works that objectively show and quantify the elasticity (or lack of it) of those fluids. Based on that, the aim of this study is to accurately characterise the rheology of a number of vitreous substitutes commonly used in eye surgery, use that data to better understand the flow behaviour of those fluids during saccadic movements, and assess the differences between their behaviour and that of the real VH (presented in \cite{Silva2020}). The remainder of the paper is organised as follows: the next section presents the experimental and numerical methodologies followed; the results are then presented and discussed; and the main conclusions obtained close this paper.

\section{Methodology}

\subsection{Experimental}

\subsubsection{Fluid characterisation}
Five different pharmacological fluids used in eye surgery were used in this study, including: three Silicone oils (SiO), RS-Oil 1000 from Alchimia (Padova, Italy), Siluron 2000 and Siluron 5000 both from Fluoron (Ulm, Germany); Densiron 68 from Fluoron (Ulm, Germany), which is a mixture of $70\%$ of SiO with a viscosity of approximately $5 \text{ Pa s}$ with $30\%$ of perfluorohexyloctane; and one Perfluorocarbons (PFLC), fluorinated perfluorodecalin (HPF10) from Alchimia (Padova, Italy). Note that all the SiOs used are composed of polymer chains of the same MW, except Siluron 2000 that is composed of $95\%$ of short molecular chains and $5\%$ of ultra-long molecular chains with a viscosity around $2500 \text{ Pa s}$. 

Shear measurements were performed using a DHR-2 rotational rheometer (TA Instruments) using a cone-and-plate configuration with 60 mm diameter and a $1^{\circ}$ angle.  

The extensional properties of the samples were measured using an in-house miniaturised filament breakup device \citep{Sousa2017,Sousa2018}, where the samples were subjected to the slow retraction method (SRM) \citep{Campo-Deano2010}. The SRM technique allows to investigate filament thinning of fluids in extensional flow with low viscosity and/or very low relaxation times, applying a slow plate separation (instead of a step strain deformation), which allows positioning the plates close to the critical point when a stable liquid bridge still exists \citep{Campo-Deano2010,Vadillo2012}. The extensional rheology of the pharmacological fluids was measured with cylindrical
plates with diameter $D_p=2 \text{ mm}$ and an initial gap of $L_0=0.5 \text{ mm}$ for the silicone oils, and $L_0=0.25 \text{ mm}$ for the PFLC. The SRM measurements were performed with a plate separation velocity of 5 $\mu$m/s at temperatures between $18.6\text{ }^{\circ}\text{C}$ and $23.3\text{ }^{\circ}\text{C}$. During the experiments, the evolution of the thinning fluid filament formed between the plates was monitored with video imaging using a high-speed CMOS camera (FASTCAM MINI UX100, Photron). The camera was connected to a magnifying optical lens (H) (OPTEM Zoom 70XL) allowing a variable magnification from $1\times \text{ to } 5.5\times$. The video images were afterwards analysed in Matlab (MathWorks, version R2014b), in order to identify when the filament exhibits a quasi-cylindrical shape and obtain the filament diameter as a function of time. This allows subsequently determining the relaxation time for viscoelastic fluids and the filament breakup time.

%All the images were recorded with a resulting magnification of $0.86 \text{ } \mu\text{m/pixel}$.

Surface tension measurements were performed with a drop shape analyser (model DSA25,
Kr$\ddot{\text{u}}$ss) using the pendant drop method. The measurements were performed at a temperature of ${T=20\pm2 \text{ } ^{\circ}\text{C}}$. 

\subsection{Computational Fluid Dynamics}

\subsubsection{Eye geometry and mesh}

\begin{figure}[t]
	\centering
		\subfloat[]{\includegraphics[width=0.30\textwidth]{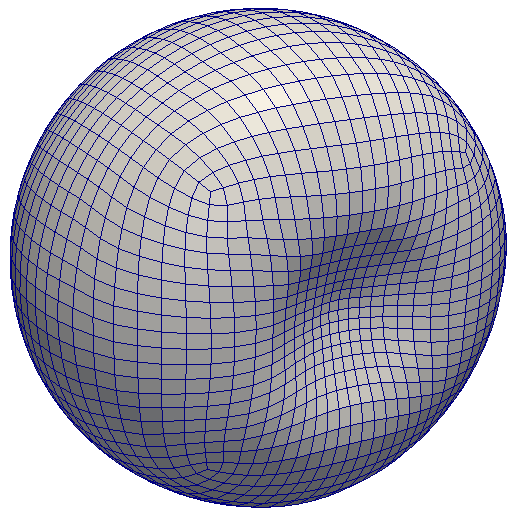}\label{fig:mesh2_a}}
		\subfloat[]{\includegraphics[width=0.265\textwidth]{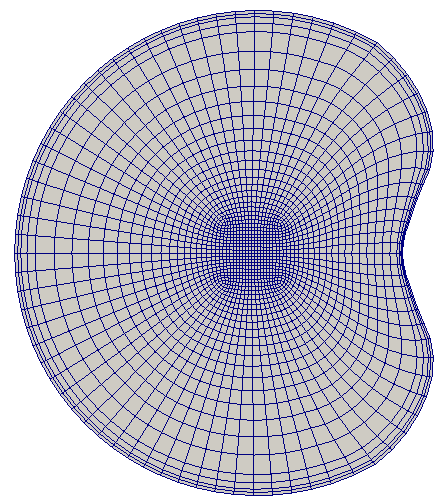}\label{fig:mesh2_refined_walls_b}}\\
		\subfloat[]{\includegraphics[width=0.30\textwidth]{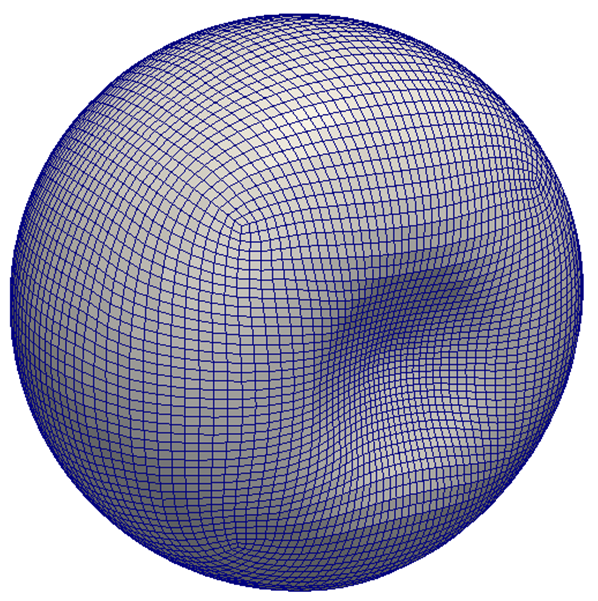}\label{fig:mesh3_a}}
		\subfloat[]{\includegraphics[width=0.25\textwidth]{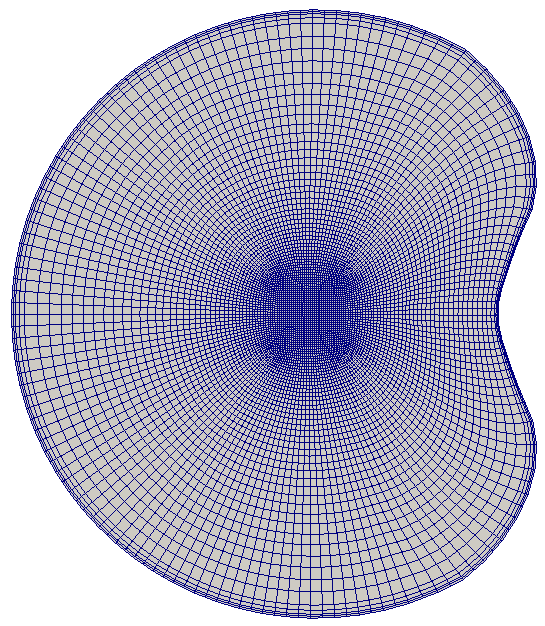}\label{fig:mesh3_refined_walls_b}}\\
	\caption{Computational meshes used to simulate the Silicone oils (a,b) and the Perfluorocarbon (c,d) flow behaviour, with (a,c) showing the surface mesh , and (b,d) the mesh at plane $z=0$.}
	\label{fig:mesh_eye}
\end{figure}

The eye geometry in this work is the same used in our previous work and detailed information can be found in \cite{Silva2020} and in the Supplementary Material 1.  
Different levels of mesh refinement and different time steps were tested to ensure the accuracy of the CFD results. For mesh purposes, the eye model was split in seven different blocks: a cube in the middle and six blocks around the central cube. Meshes with $10^3$, $20^3$, $40^3$ and $80^3$ cells per block, with additional refinements in the cell adjacent to the wall (region with higher gradients of velocity) were tested. The velocity profiles, in the vitreous cavity, at different times of a saccadic movement with amplitude $A=40^{\circ}$ were compared and for the Silicone Oils no significant differences were found between the results obtained with the meshes with $20^3$ and $40^3$ cells per block, while for the HPF10 the meshes with $40^3$ and $80^3$ cells per block showed no significant differences. Based on these results, meshes with $20^3$ and $40^3$ cells per block were used to simulate the flow behaviour of the Silicone oils and the Perfluorocarbon, respectively (see Fig. \ref{fig:mesh_eye}). Two time steps were tested ($\Delta t=10^{-4} \text{ s}$ and $10^{-5} \text{ s}$), with the obtained velocity profiles showing no meaningful differences. The time step of $\Delta t=10^{-4} \text{ s}$ was thus used in all the remaining simulations.

\subsubsection{Saccadic movements}

Saccadic movements with angular displacements, $A$, between 10 and $40^{\circ}$ were computed using the fifth order polynomial equation presented by \cite{Repetto2005}, based on the experimental measurements of \cite{Becker1989}. Detailed information can be found in our previous work \citep{Silva2020} and in Supplementary Material 2. Those details are not repeated here for briefness.

\subsubsection{Governing equations}

During this work the whole eye model moves as a solid body, with a 
prescribed motion. In order to include the effect of the dynamic mesh, the convective terms in the governing equations needed to be corrected with the grid velocity.
In this moving mesh framework, the continuity equation can be written as 

\begin{equation}
\triangledown \cdot \textbf{u}=0
\end{equation}

where \textbf{$\textbf{u}$} is the velocity vector. 

The momentum equation can be written as

\begin{equation}
\rho \bigg[\frac{\partial \textbf{u}}{\partial t}+ \triangledown \cdot \textbf{u} (\textbf{u}-\textbf{u}_g)\bigg]= - \triangledown p+\triangledown \cdot \bm{\tau} 
\end{equation}

where $\rho$ is the fluid density, $p$ the pressure, $t$ is time, $\bm{\tau}$ is the extra-stress tensor, and $\textbf{u}_g$ the velocity at which the grid surface is moving. Besides, the space conservation law (SCL) also needs to be taken into account for moving grids,
\begin{equation}
\frac{\text{d}}{\text{d}t}  \int_V \text{d}V - \int_S \textbf{u}_g \cdot \textbf{n}\text{d}S = 0
\end{equation}

where $V$ and $S$ are the volume and surface of the grid.

%To solve the momentum equation, the extra-stress tensor $\bm{\tau}$ constitutive law has to be calculated. 
The fluids HPF10, Densiron 68, RS-Oil 1000 and Siluron 5000  were simulated as Newtonian fluids in accordance with the rheology measurements presented in Section \ref{rheo_res}. The Newtonian stress tensor $\bm{\tau}_{s}$ is given by 

\begin{equation}
\bm{\tau}_{s} = \eta_s(\triangledown \textbf{u} + \triangledown \textbf{u}^{\text{T}})
\label{eq:newt_solv_cont}
\end{equation}

where $\eta_s$ is the solvent viscosity. 
The Siluron 2000 fluid exhibits a weakly elastic behaviour. To assess if such a low value of elasticity affects the fluid flow when compared with the Newtonian fluid behaviour, Siluron 2000 was simulated as a Newtonian fluid or as a viscoelastic fluid using an Oldroyd-B model. This viscoelastic model is frequently used in the simulation of Boger fluids, and predicts a constant shear viscosity ($\eta_0=\eta_{s}+\eta_{p}$, where $\eta_{p}$ is the polymer viscosity) and a quadratic increase of the first normal-stress difference with shear rate in a steady viscometric flow \citep{Morrison2001}. For Siluron 2000, the flow history is important and the total extra-stress tensor is split in a polymeric contribution $\bm{\tau}_{p}$ and a solvent contribution $\bm{\tau}_{s}$: $\bm{\tau}=\bm{\tau}_{s} + \bm{\tau}_{p}$. The polymer contribution $\bm{\tau}_{p}$ is solved based on the upper-convected Maxwell constitutive equation \citep{Owens2002}

\begin{equation}
\bm{\tau}_{p} + \lambda\overset{\triangledown}{\bm{\tau}_{p}} = \eta_{p}(\triangledown \textbf{u} + \triangledown \textbf{u}^{T})
\end{equation} 

where $\lambda$ is the relaxation time and
$\overset{\triangledown}{\bm{\tau}_{p}}$ is the upper-convected time derivative of $\bm{\tau}_{p}$, defined as 
$\overset{\triangledown}{\bm{\tau}_{p}}=\frac{\partial \bm{\tau}_{p}}{\partial t}+\triangledown\cdot\left[{\bm{\tau}_{p}}\left(\textbf{u}-\textbf{u}_{g}\right)\right]-{\bm{\tau}_{p}}\cdot\triangledown\textbf{u}-\triangledown\textbf{u}^{\text{T}}\cdot{\bm{\tau}_{p}}$ \citep{Pimenta2017}.

For Siluron 2000, two different solvent viscosities ratios were tested: $\beta=\eta_s/(\eta_{s}+\eta_{p})=\eta_s/\eta_0=0.98$ and $0.95$. The viscosity values used in the simulations for all the SiOs are presented in Table \ref{tab:oldroydB}.  

\begin{table}[b]
	\centering
	\caption{Viscosity values for the Newtonian and the Oldroyd-B models used in the CFD simulations.}
	\label{tab:oldroydB}
	\begin{adjustbox}{width=0.7\textwidth}
		\begin{tabular}{ccccccc}
			\hline
			\multirow{3}{*}{Fluids} & \begin{tabular}[c]{@{}c@{}}Newtonian\\  model\end{tabular} &  \multicolumn{5}{c}{\begin{tabular}[c]{@{}c@{}}Oldroyd-B\\  model\end{tabular}}                                                                         \\ \cline{2-7} 
			& \multirow{2}{*}{$\eta$ (Pa s)} & \multicolumn{2}{c}{$\beta=0.98$}                       & \multicolumn{2}{c}{$\beta=0.95$}                      & \multirow{2}{*}{$\lambda$ (s)} \\
			&                                & $\eta_p$ (Pa s) & \multicolumn{1}{c}{$\eta_s$ (Pa s)} & $\eta_p$ (Pa s) & \multicolumn{1}{c}{$\eta_s$ (Pa s)} &                                \\ \hline
			\multicolumn{1}{l}{HPF10}                                 & \multicolumn{1}{c}{$3.57 \times 10^{-3}$ }                           & --          & \multicolumn{1}{c}{--}          & --          & \multicolumn{1}{c}{--}          & --                         \\
			\multicolumn{1}{l}{RS-Oil 1000}                                 & \multicolumn{1}{c}{0.73}                           & --          & \multicolumn{1}{c}{--}          & --          & \multicolumn{1}{c}{--}          & --                         \\
			\multicolumn{1}{l}{Densiron 68}                                & \multicolumn{1}{c}{1.07}                           & --          & \multicolumn{1}{c}{--}          & --          & \multicolumn{1}{c}{--}          & --                         \\
			\multicolumn{1}{l}{Siluron 2000}                              & \multicolumn{1}{c}{1.71}                           & 0.0342          & \multicolumn{1}{c}{1.6758}   & 0.0855          & \multicolumn{1}{c}{1.6245}          & 0.0068                         \\
			\multicolumn{1}{l}{Siluron 5000}                                 & \multicolumn{1}{c}{4.57}                           & --          & \multicolumn{1}{c}{--}        & --         & \multicolumn{1}{c}{--}          & --                         \\ \cline{1-7}
		\end{tabular}
	\end{adjustbox}
\end{table}

For the fluids and flow conditions under study, the maximum Reynolds number reached in the simulations can be calculated as $\text{Re} = \frac{\rho U_{max} R}{\eta_0}$, where $R$ is the eye radius, $U_{max}$ is the maximum velocity reached for each degree of movement, and $\eta_0$ is the zero shear viscosity of the fluid. The maximum Reynolds number reached for the SiOs fluids is below $\text{Re}=2$, while the maximum value for HPF10 is Re$\approx717$.

\subsubsection{Numerical method}

The RheoTool package \citep{Pimenta2017} in the open-source finite-volume code OpenFOAM® (version 2.2.2) was used to solve the governing equations. The no-slip boundary condition was set in the whole eye surface. The time-derivatives were discretised using a second-order implicit Backward Euler scheme, and the semi-implicit method for pressure-linked equations-Consistent (SIMPLEC) algorithm was selected to ensure the velocity-pressure coupling in a segregated way. The convective  terms of both the momentum and constitutive equations were discretised with the Convergent and Universally Bounded Interpolation Scheme for the Treatment of Advection (CUBISTA) \citep{Alves2003}. A zero pressure gradient normal to the walls was assumed and the polymeric stress tensor was extrapolated from the cells adjacent to the wall \citep{Pimenta2017}. Finally, the log-conformation tensor formulation was used to solve the constitutive equation.

\section{Results and discussion}

In this section, first we present the rheological characterisation of the pharmacological fluids used in eye surgery. Then, we investigate numerically the effect of elasticity on the velocity and wall shear stress (WSS) of Siluron 2000. Finally, we compare and discuss the behaviour of the various vitreous substitutes during saccadic eye movements.

\subsection{Rheological characterisation}
\label{rheo_res}

\begin{figure}[]
	\centering
	{\includegraphics[scale=0.25]{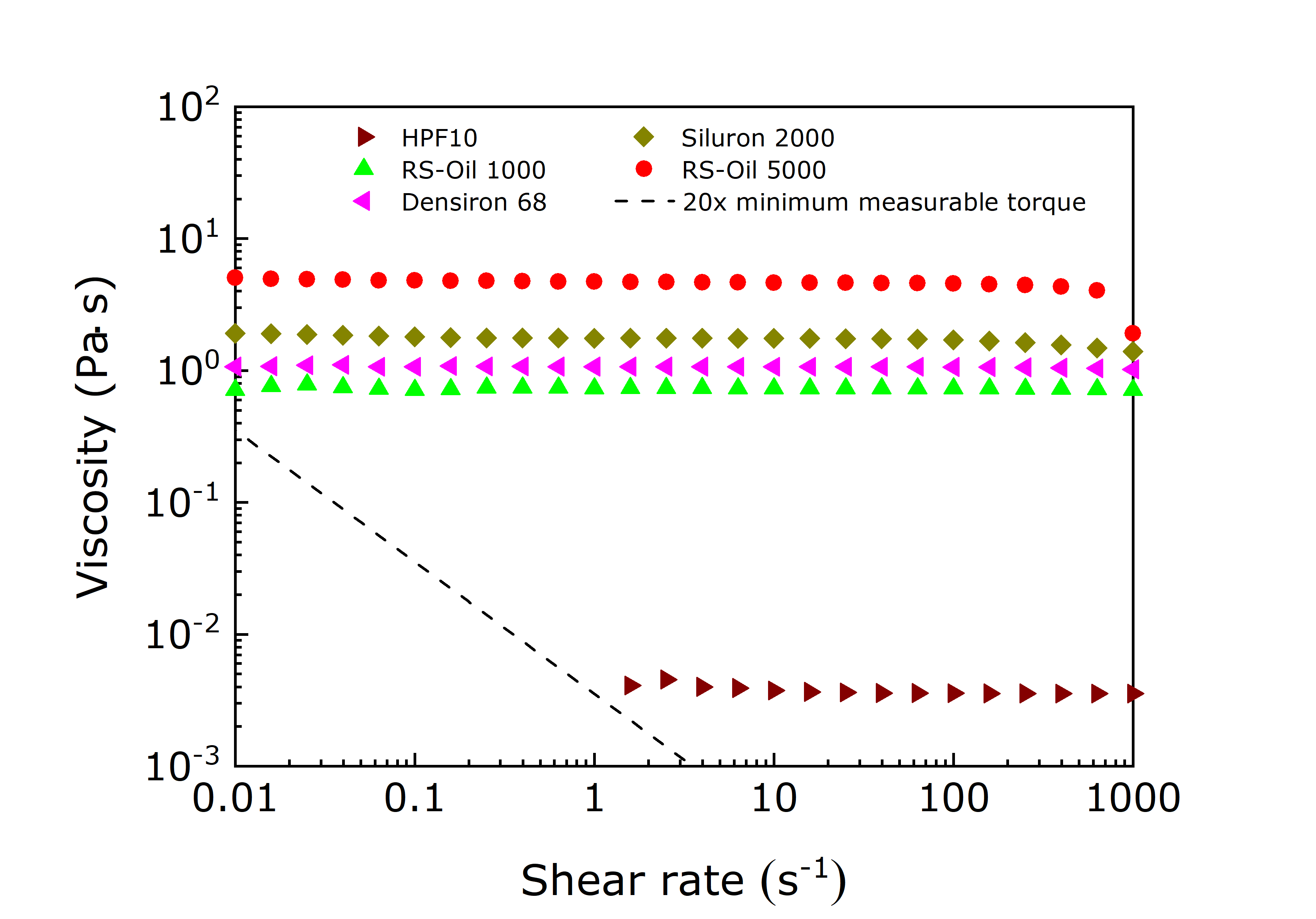}}
	\caption{Flow curve of VH substitutes, measured at $T=37 \text{ }^{\circ}\text{C}$. The dashed black line correspond to 20x the minimum measurable torque.}
	\vspace{0em}
	\label{fig:flow_curve_pf}
\end{figure}

The flow curves for the fluids under study are shown in Fig. \ref{fig:flow_curve_pf} for a constant temperature of $T=37 \text{ } ^{\circ}\text{C}$. All the pharmacological fluids show a constant viscosity over the range of shear rates tested. 

\begin{figure}[]
	\centering
	{\includegraphics[width=0.4\textwidth]{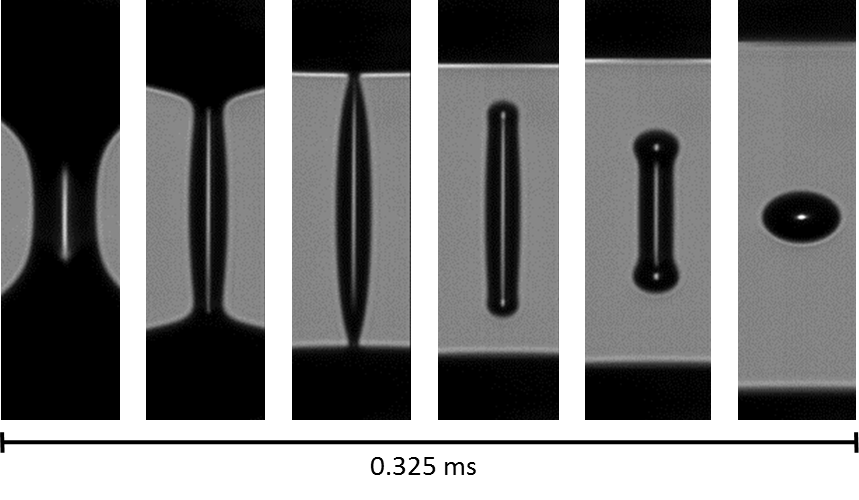}} %\subcaption{(a) Densiron 68}}
	\caption{Filament thinning dynamics induced by the SRM for the HPF10 fluid. 
		The measurement was performed at $T=21\pm2 \text{ }^{\circ}\text{C}$, the plates 
		were moved with a constant velocity of 5 $\mu$m/s and the plate diameter is $D_p$= 2 mm.}
	\label{fig:PFLC} 
\end{figure}

When subjected to an extensional flow with a plate separation with a velocity of $5 \text{ } \mu \text{m/s}$, the HPF10 filament breakup occurs close to the plates (see Fig. \ref{fig:PFLC}), as is typical behaviour of low viscosity Newtonian fluids. For the remaining fluids, the breakup occurs close to the centre of the filament due to their higher viscosity. For a Newtonian fluid the thread thins linearly in time with the midpoint radius decreasing as \citep{Papageorgiou1995}  

\begin{equation}
R_{min}=0.0709 \frac{\sigma}{\eta}(t_0-t)
\label{eq:Papageorgiou}
\end{equation}

where $\sigma$ is the surface tension, $\eta$ the shear viscosity of the fluid and $t_0$ the time at filament breakup.

\begin{figure}[]
	\centering
	\subfloat[]{\includegraphics[width=0.4\textwidth]{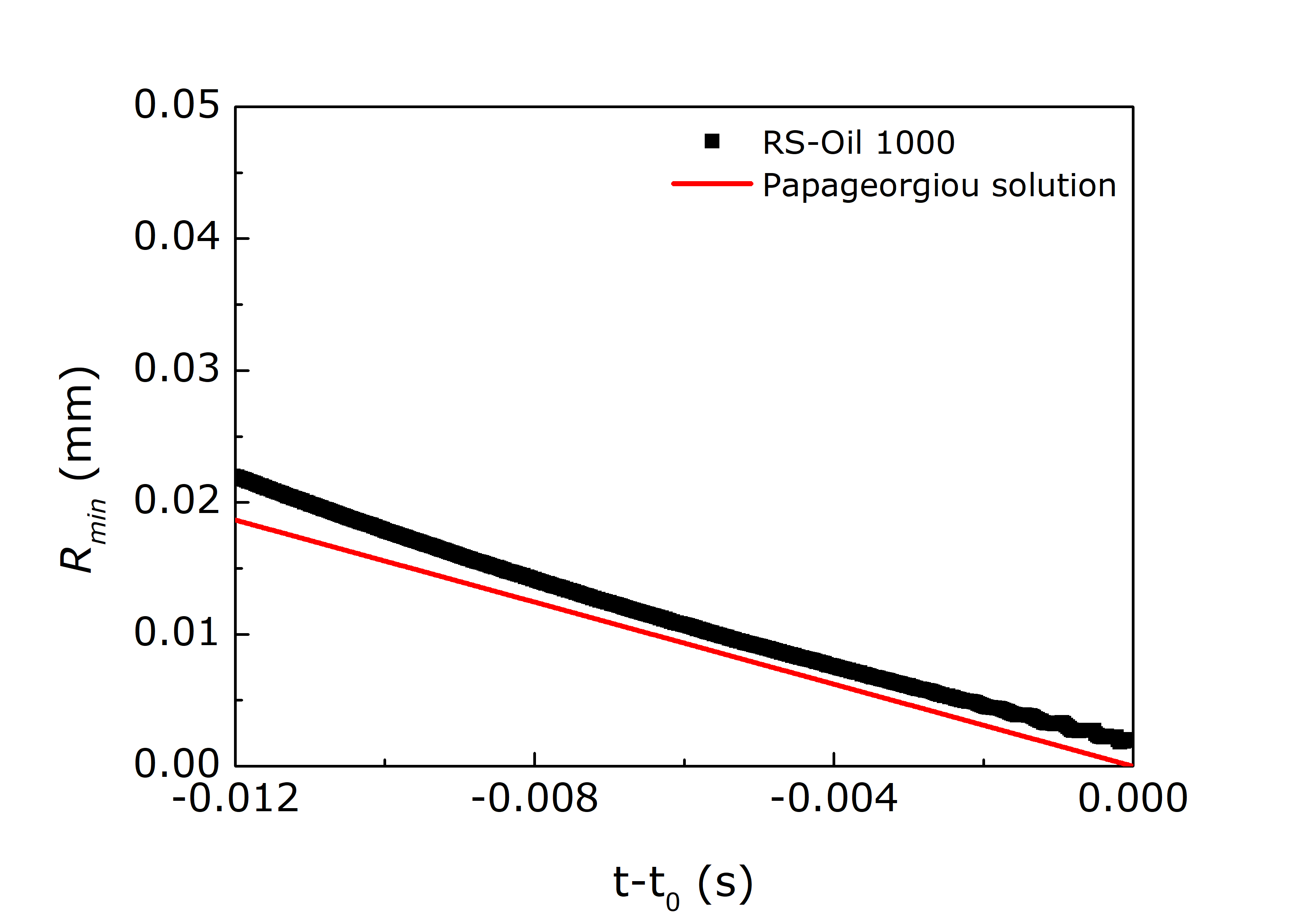}} %\subcaption{(b) RS-Oil 1000}}
	\subfloat[]{\includegraphics[width=0.4\textwidth]{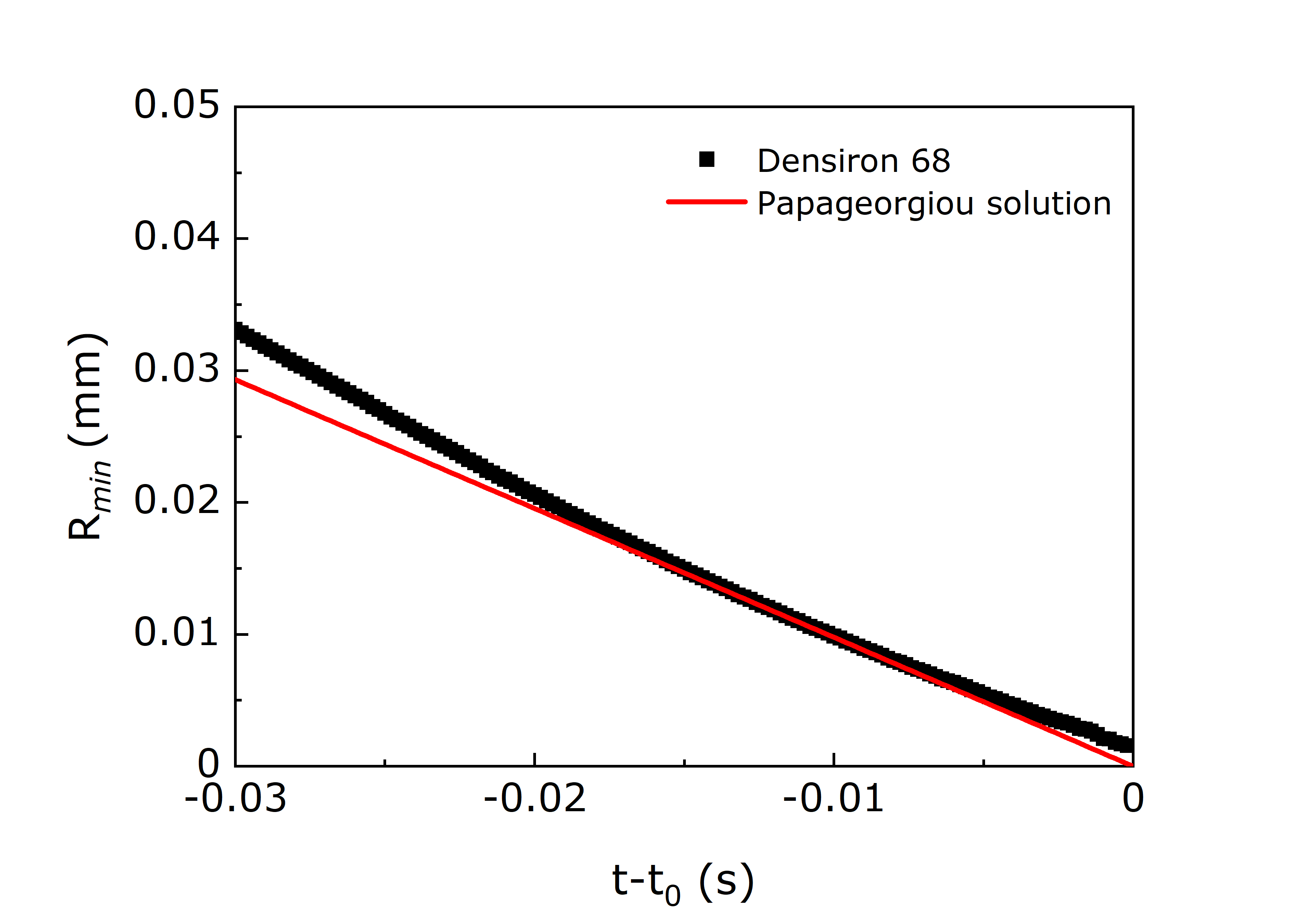}}\\%\subcaption{(a) Densiron 68}}
	\subfloat[]{\includegraphics[width=0.4\textwidth]{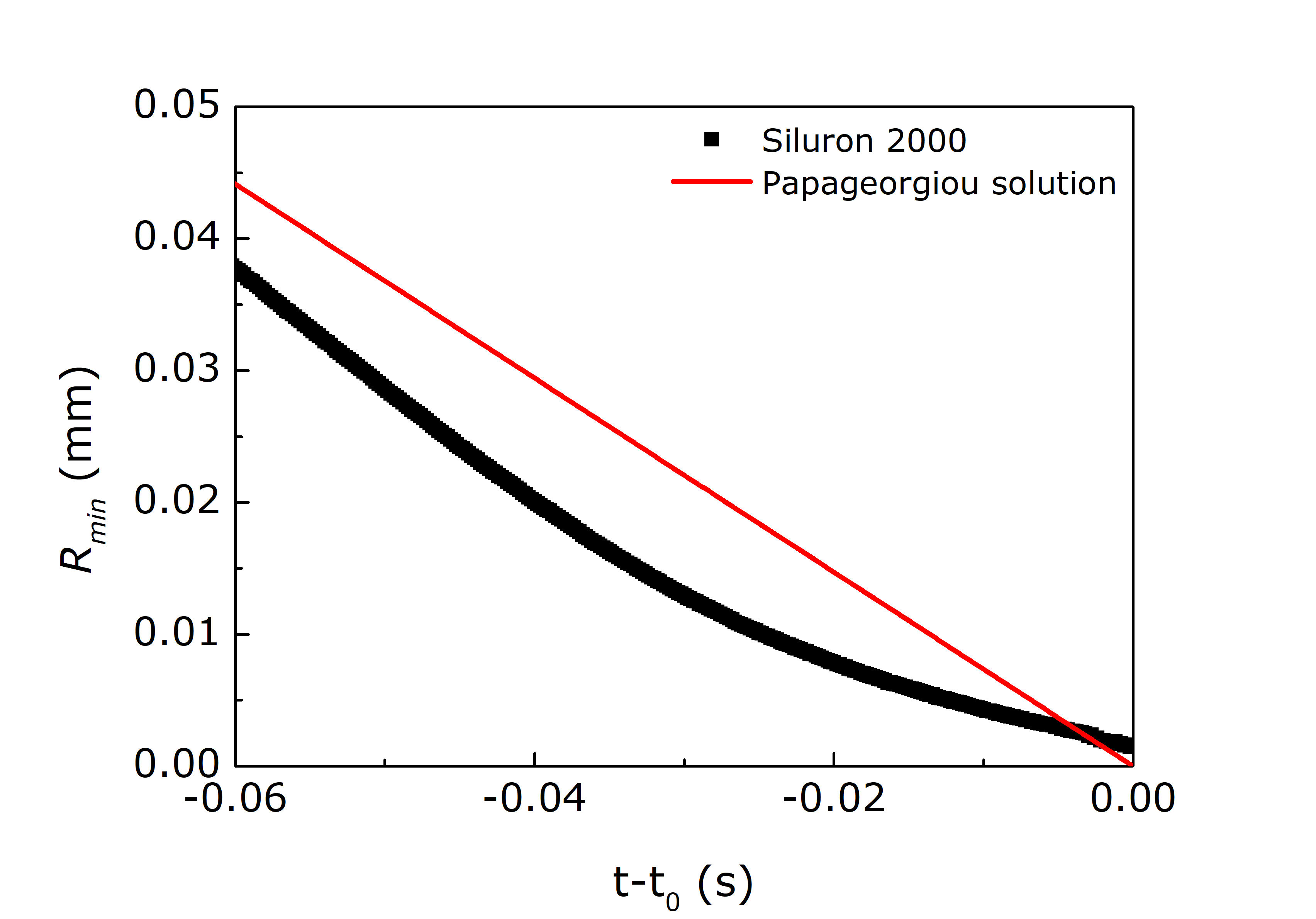}}%\subcaption{(c) Siluron 2000}} 
	\subfloat[]{\includegraphics[width=0.4\textwidth]{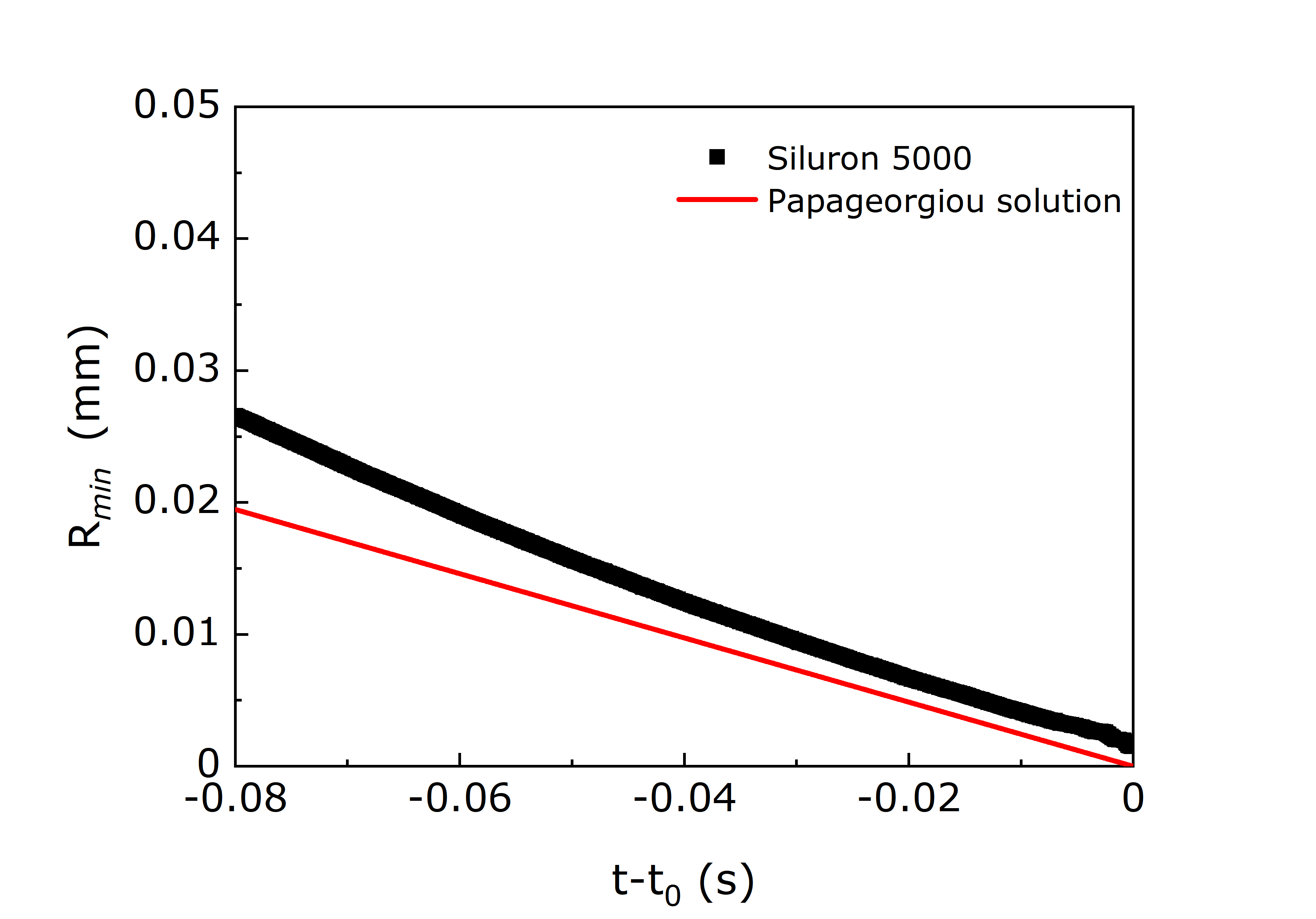}}\\%\subcaption{(d) Siluron 5000}}
	\caption{Time evolution of the minimum radius ($R_{min}$) close to breakup when a quasi cylindrical shape is observed and the Papageorgiou solution \citep{Papageorgiou1995}. $t_0$ is the time of the filament breakup. All the measurements were performed at $T=21\pm2 \text{ }^{\circ}\text{C}$, with a plate with diameter 
		$D_p = 2 \text{ mm}$ and an initial gap of $0.5 \text{ mm}$.}
	\label{fig:microcaber_papageorgiou} 
\end{figure}

Based on the shear viscosity and the surface tension values measured experimentally (see Table \ref{tab:sur_tension_pf}), Eq. \ref{eq:Papageorgiou} was tested for all SiO based fluids. A comparison between the miminum radius of the filament $R_{min}$ before the breakup for all the fluids studied and the theoretical behaviour predicted by \cite{Papageorgiou1995} is presented in Fig. \ref{fig:microcaber_papageorgiou}. 

\begin{table}[]
	\centering
	\caption{Surface tension of the pharmacological fluids analysed in this study, at $T=21\pm1 \text{ }^{\circ}\text{C}$.}
	\label{tab:sur_tension_pf}
	\begin{tabular}{ccc}
		\hline
		\multicolumn{1}{c}{Fluid}         & \multicolumn{1}{c}{\begin{tabular}[c]{@{}c@{}}Average surface tension \\ (mN/m)\end{tabular}} & \multicolumn{1}{c}{\begin{tabular}[c]{@{}c@{}}Standard deviation \\ (mN/m)\end{tabular}} \\ \hline
		HPF10                             & 20.5                                                                                         & 0.4                                                                                     \\ 
		Densiron 68                       & 20.8                                                                                         & 0.2                                                                                     \\
		RS-Oil 1000                       & 21.9                                                                                         & 0.3                                                                                    \\
		Siluron 2000                      & 22.0                                                                                         & 0.3                                                                                    \\
		Siluron 5000                      & 21.7                                                                                         & 0.6                                                                                     \\ \hline
	\end{tabular}
\end{table}

\begin{figure}[]
	\centering
	{\includegraphics[scale=0.25]{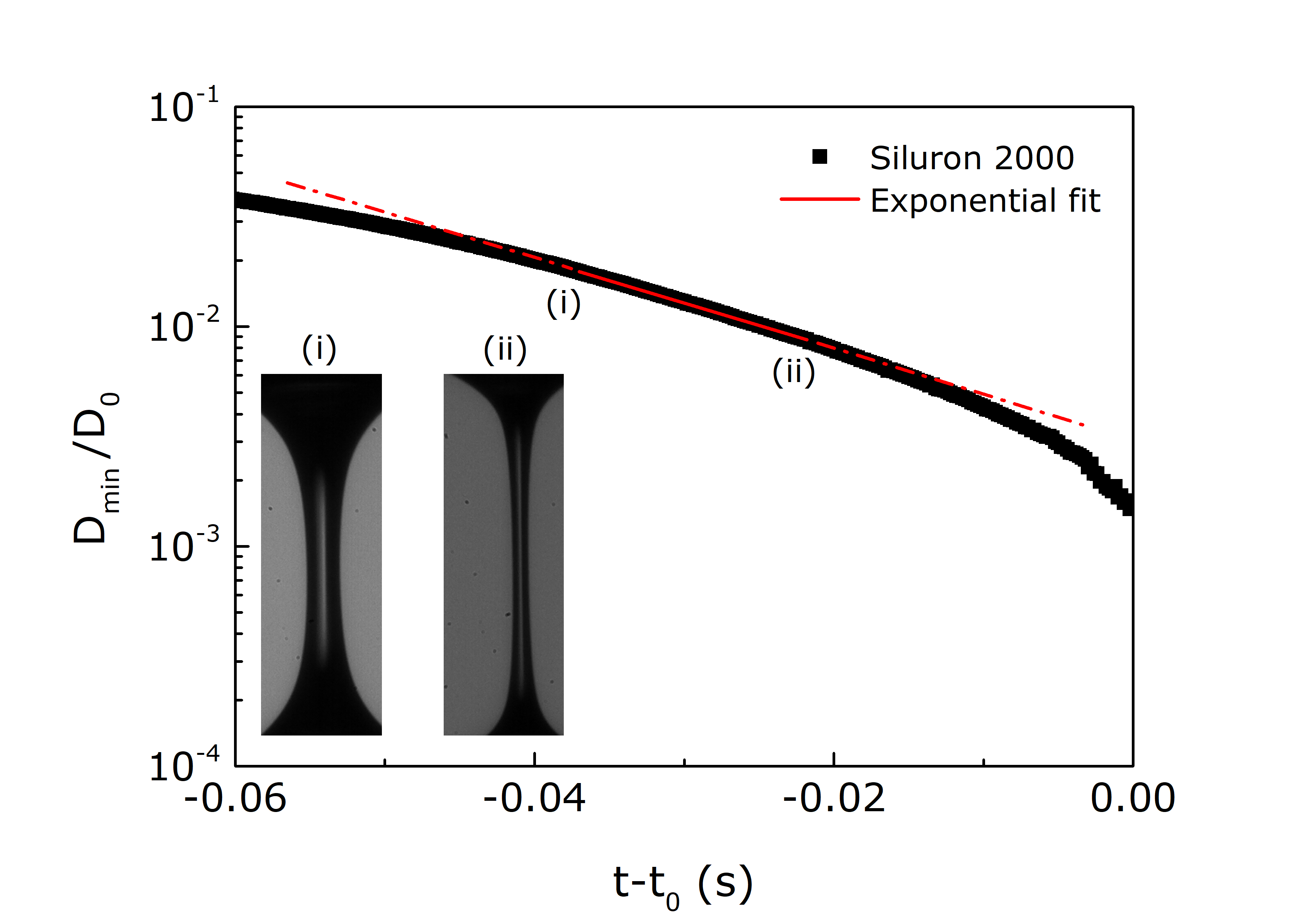}}
	\caption{Time evolution of the minimum filament diameter ($D_{min}/D_0$) close to breakup and the exponential fit used to calculate the extensional relaxation time $\lambda$ for 
		Siluron 2000 fluid. $t_0$ is the time of the filament breakup. (i) and (ii) show the shape of the filament for the first and last points, respectively, of the exponential region used to calculate the relaxation time, where a cylindrical section of the filament is clearly visible.
		The measurement was performed at $T=21\pm2 \text{ }^{\circ}\text{C}$, with a plate with diameter 
		$D_p = 2 \text{ mm}$ and an initial gap of $0.5 \text{ mm}$.}
	\label{fig:microcaber} 
\end{figure}

Note that the Papageorgiou's solution is calculated based on the shear rheology and average surface tension measured experimentally. As a consequence, the accuracy of the Papageorgiou solution depends on those measurements and slight differences may occur. The results presented in Fig. \ref{fig:microcaber_papageorgiou} show that close to the filament breakup, the diameter decay with time for the RS-Oil 1000, Densiron 68 and Siluron 5000 fluids is linear, with a slope similar to the Papageorgiou solution, indicating that those fluids behave as Newtonian fluids. However, the Siluron 2000 behaviour does not follow the Papageorgiou's solution, because there is a distinct slope between the experiments and the theoretical Newtonian profile. This fluid shows elastic behaviour and it is possible to calculate an extensional relaxation time. Figure \ref{fig:microcaber} shows the evolution of the minimum filament thinning diameter ($D_{min}/D_0$) for Siluron 2000 fluid close to breakup, and the exponential fit $D_{min}/D_0 \propto e^{-t/3\lambda}$ performed to obtain the relaxation time, $\lambda$ \citep{Entov1997}. Note that just one case is shown here. However, experiments were repeated three times to assess repeatibility, and the average relaxation time obtained was $\lambda=6.8 \pm 0.2 \text{ ms}$. 

In a previous work \citep{Silva2017}, we have shown that VH liquid phase viscosity decreases with the increase of shear rate (shear-thinning behaviour). The fact that these VH substitutes cannot be used as a permanent VH substitute may be related to the inability of the pharmacological fluids to mimic the shear-thinning behaviour, as they behave differently under deformation.

\subsection{Siluron 2000: assessing the impact of viscoelasticity on the flow behaviour}

\begin{figure}[!ht]
	\centering
	\subfloat[]{\includegraphics[scale=0.21]{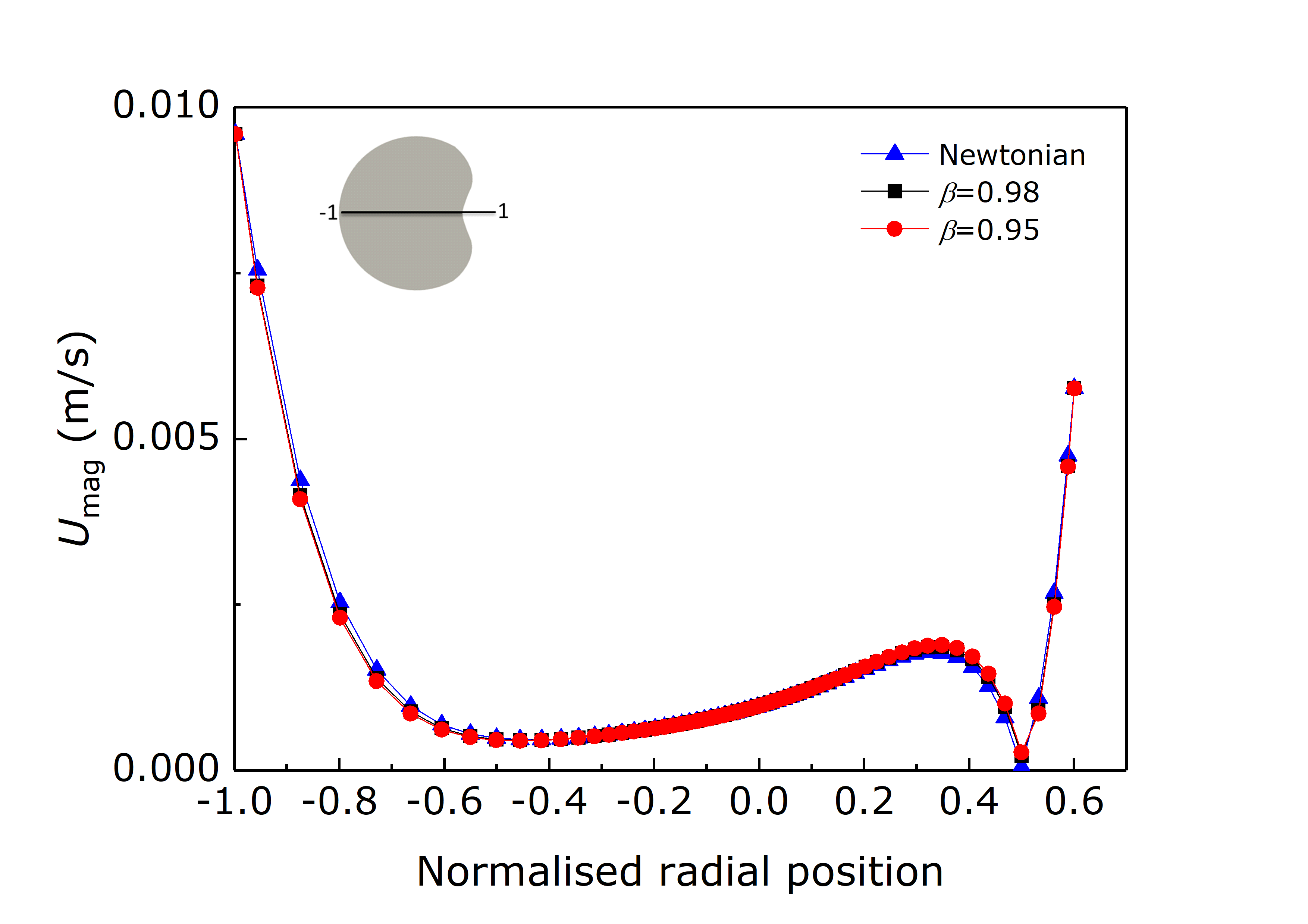}\label{fig:Newt_oldB_Siluron2000_005s}}
	\subfloat[]{\includegraphics[scale=0.21]{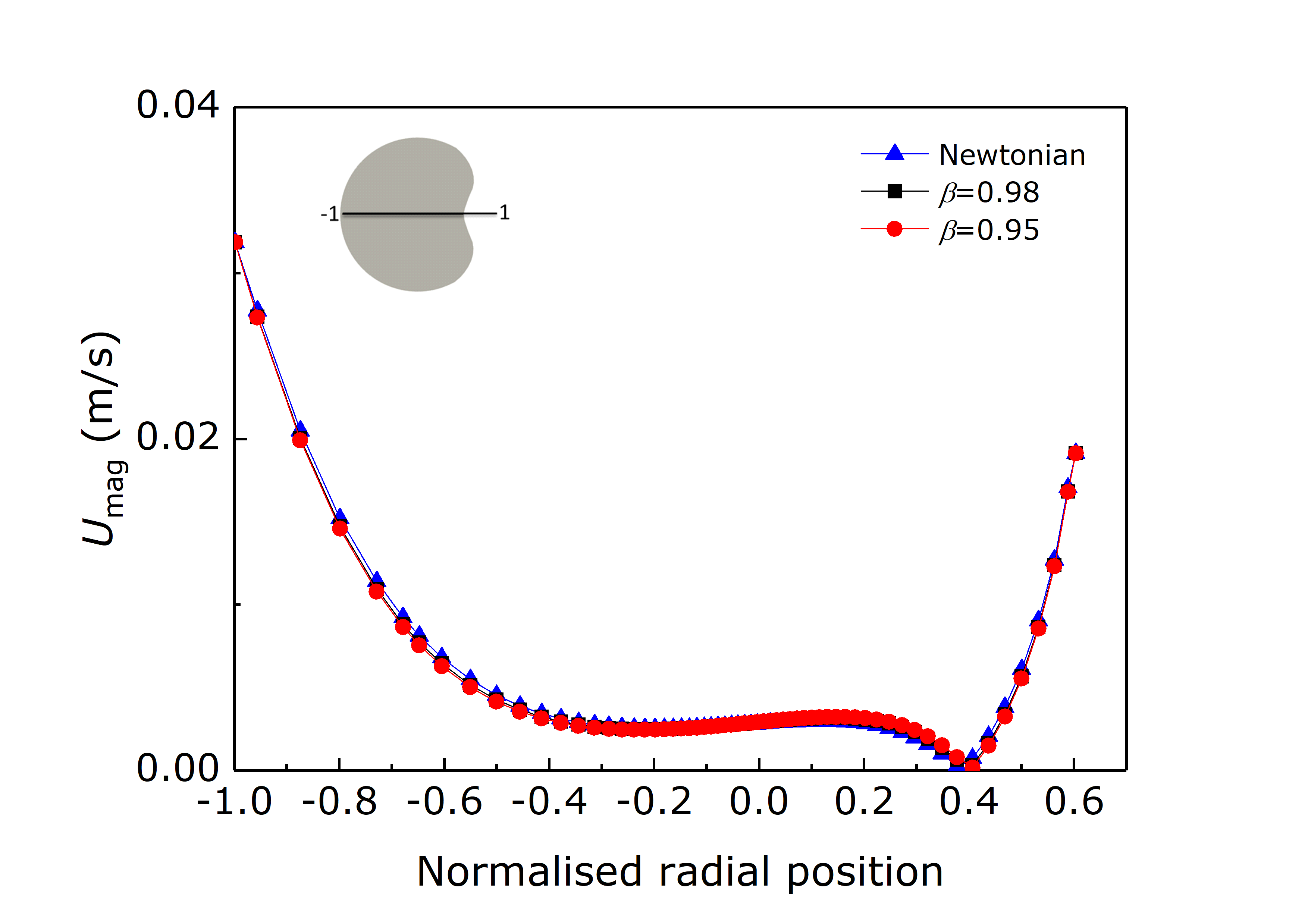}\label{fig:Newt_oldB_Siluron2000_04ts}}\\
	\subfloat[]{\includegraphics[scale=0.21]{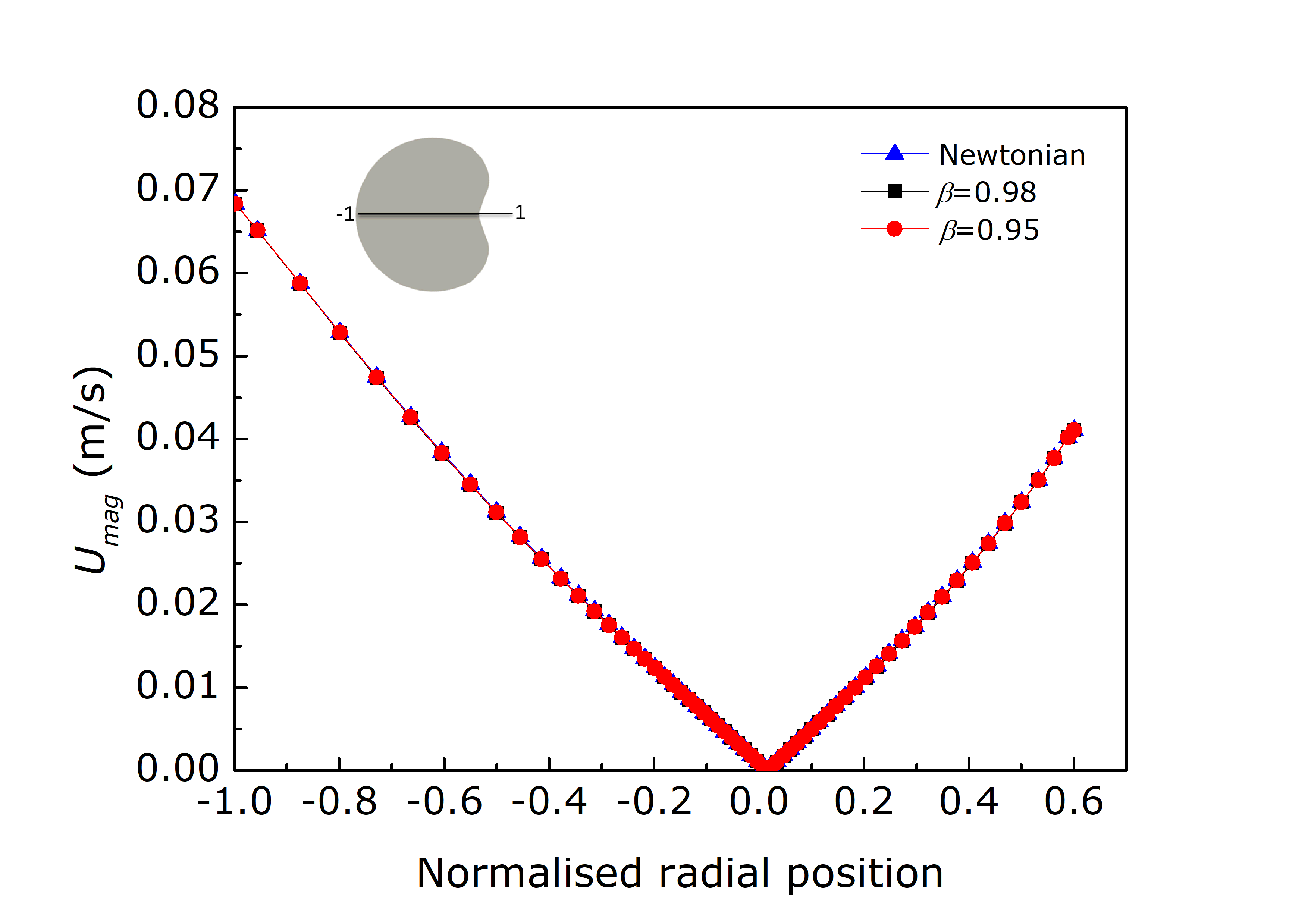}\label{fig:Newt_oldB_Siluron2000_0225s}}
	\subfloat[]{\includegraphics[scale=0.21]{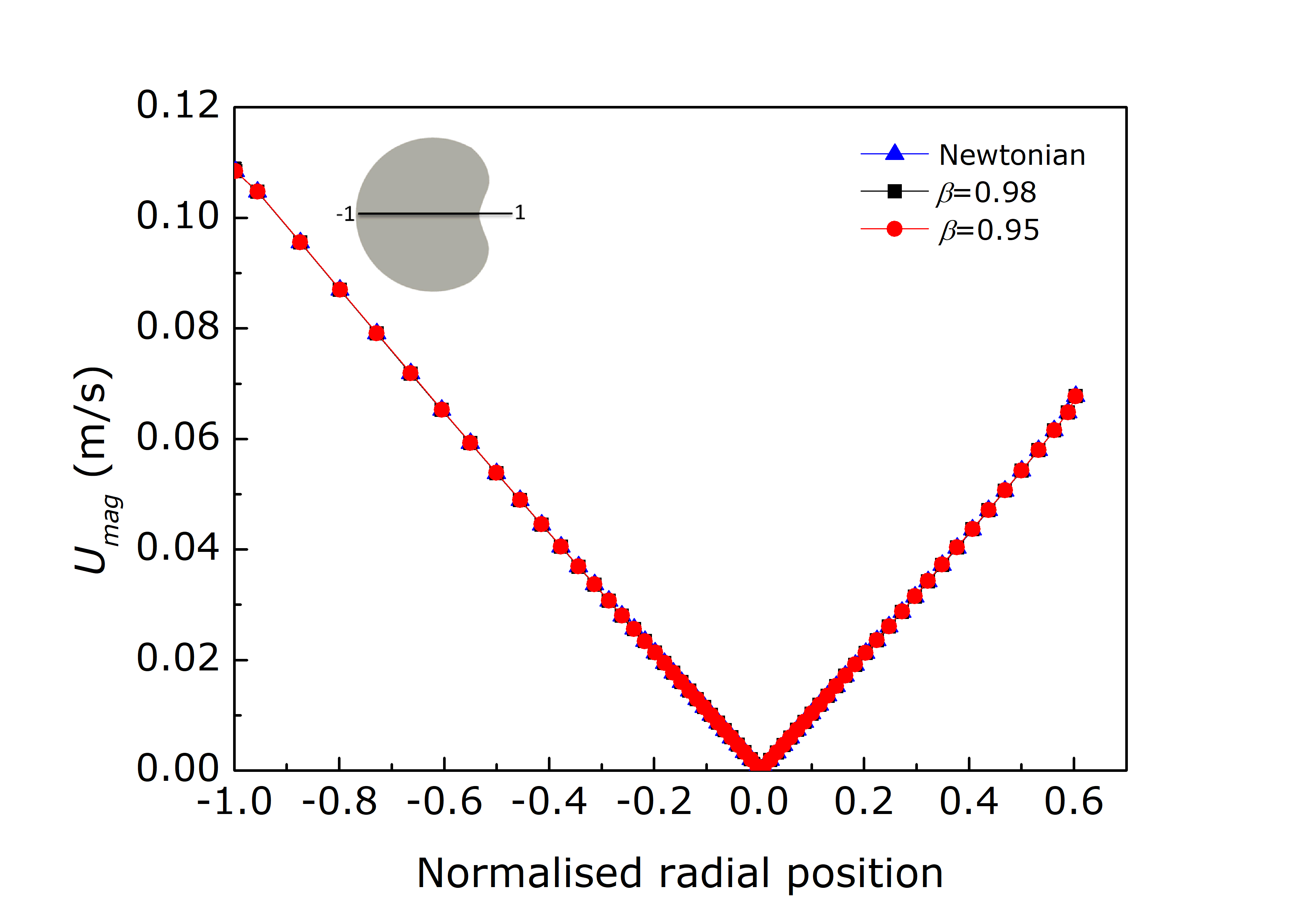}\label{fig:Newt_oldB_Siluron2000_0375s}}\\
	\subfloat[]{\includegraphics[scale=0.21]{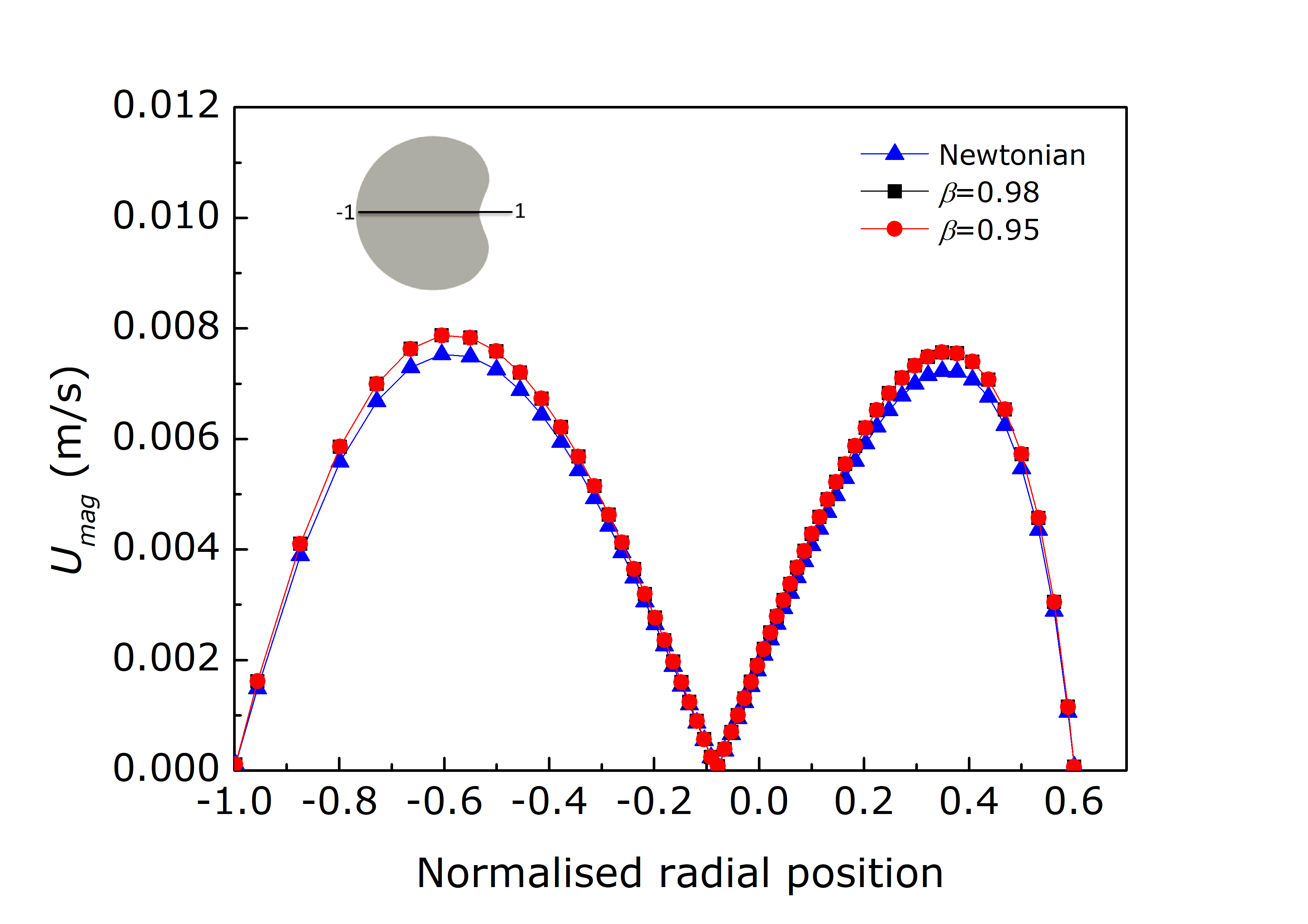}\label{fig:Newt_oldB_Siluron2000_05s}}
	\subfloat[]{\includegraphics[scale=0.21]{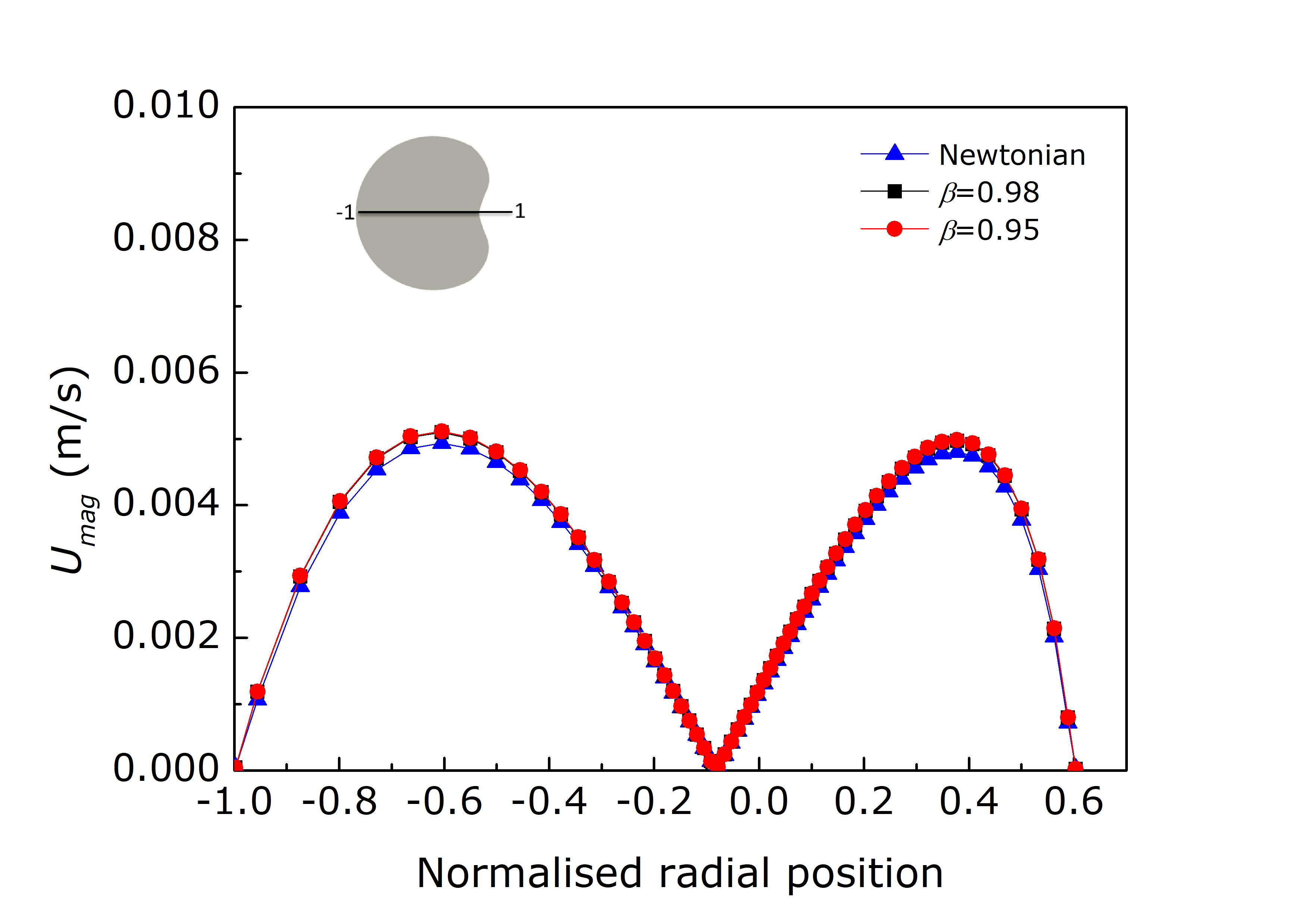}\label{fig:Newt_oldB_Siluron2000_125s}}\\
	\caption{Velocity magnitude along the normalised radial position ($x/R$) in the plane $z=0$ for Siluron 2000
		for a saccadic movement of (a,c,e) $A=10^{\circ}$ and (b,d,f)  $A=40^{\circ}$ at (a,b) $t=0.04t_D$, (c,d) $t=t_p$ and (e,f) $t=t_D$. 
		The figures compare the results obtained using a Newtonian model and an Oldroyd-B model with two different solvent viscosity ratios, $\beta=0.98$ and $\beta=0.95$.}
	\label{fig:New_Old_comparisons}
\end{figure}

To try to understand how the elasticity can affect the flow behaviour of Siluron 2000, this fluid was simulated as a Newtonian fluid and as a viscoelastic fluid using the Oldroyd-B model, with solvent viscosity ratios $\beta=0.98$ and $\beta=0.95$. The velocity magnitude in the midline at the plane $z=0$, for saccadic movements  with amplitudes $A=10^{\circ}$ and $40^{\circ}$, are shown in Fig. \ref{fig:New_Old_comparisons}. Overall, for both amplitudes, the differences observed are small and the velocity profiles follow the same trend for all fluid models considered. For both amplitudes, it is possible to observe a minor effect of the fluid elasticity for the time where the wall velocity is zero, $t=t_D$ (see Fig. \ref{fig:Newt_oldB_Siluron2000_05s} and Fig. \ref{fig:Newt_oldB_Siluron2000_125s}), in particular in the region of the local max $U_{mag}$. Additionally, for time $t=0.04t_D$ it is also possible to observe a very small difference in the velocity magnitude when the Siluron 2000 fluid is subjected to a saccadic movement with $A=10^{\circ}$ (see Fig. \ref{fig:Newt_oldB_Siluron2000_005s}). For time $t=t_p$, the velocity magnitudes are mostly the same for the Newtonian fluid and the Oldroyd-B fluids, for both amplitudes under study (see Fig. \ref{fig:Newt_oldB_Siluron2000_0225s} and Fig. \ref{fig:Newt_oldB_Siluron2000_0375s}). Finally, there are no significant differences in the velocity field for the two different solvent viscosity ratios studied ($\beta=0.98$ and $\beta=0.95$).

\begin{figure}[]
	\centering
	\subfloat[]{\includegraphics[scale=0.21]{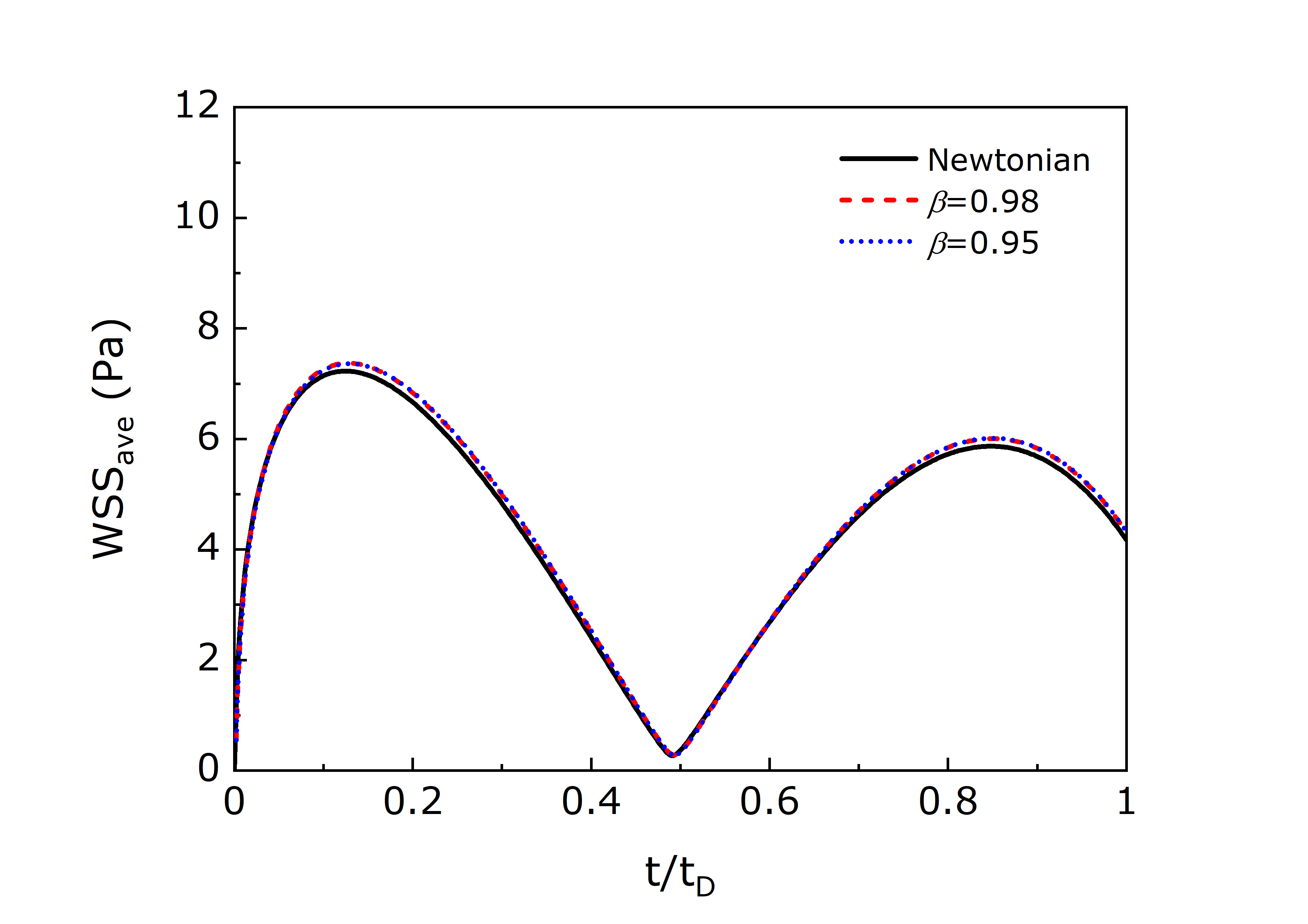}\label{fig:WSS_vs_time_Siluron2000_10degrees}}
	\subfloat[]{\includegraphics[scale=0.21]{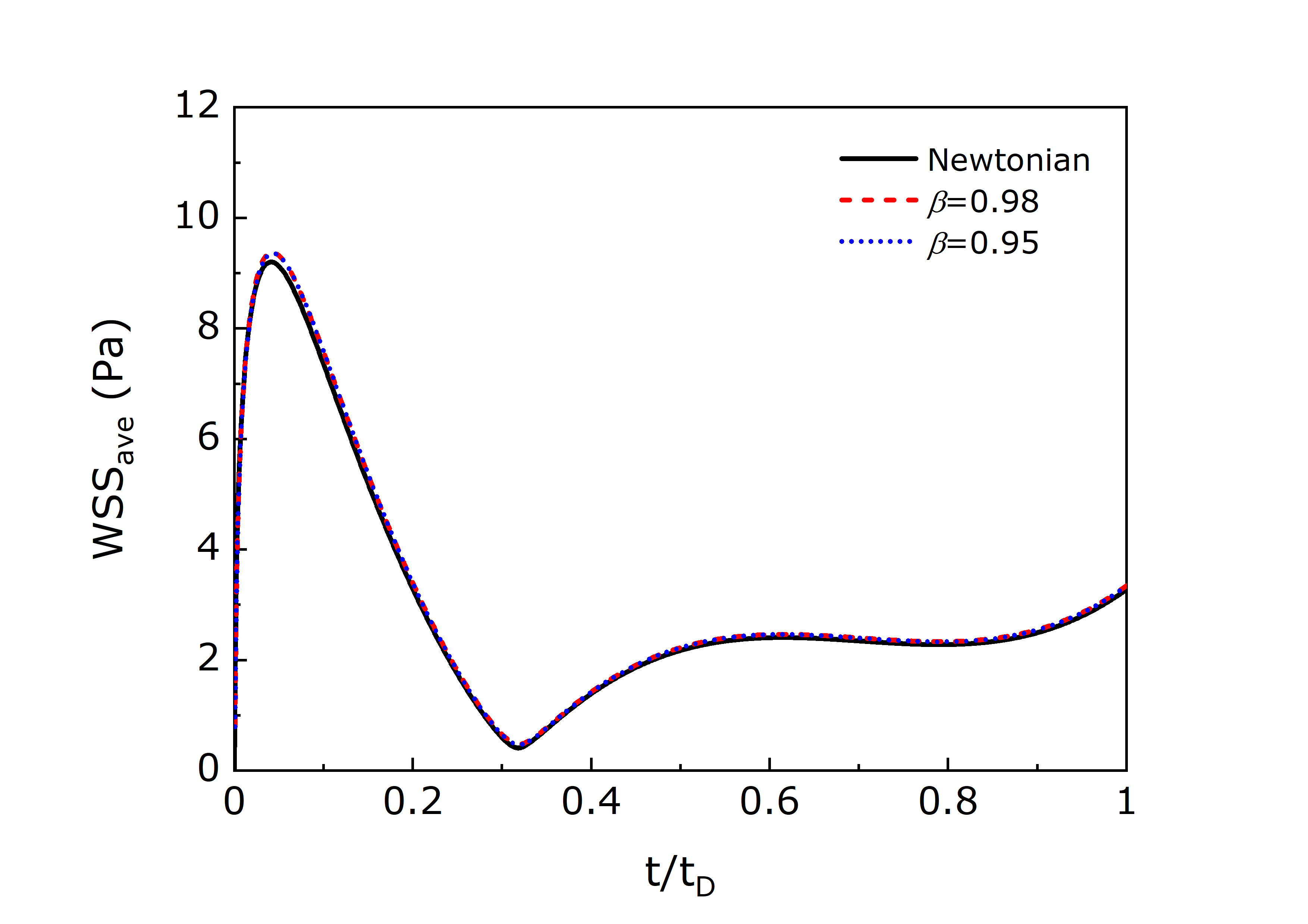}\label{fig:WSS_vs_time_Siluron2000}}\\
\caption{Time variation of the average WSS for Siluron 2000
	for saccadic movements of amplitude (a) $A=10^{\circ}$ and (b)  $A=40^{\circ}$, using a Newtonian model and an Oldroyd-B model with two different solvent viscosity ratios, $\beta=0.98$ and $\beta=0.95$.}
\label{fig:New_Old_WSScomparisons}
\end{figure}

The time evolution of the average WSS was also computed for Siluron 2000 for saccadic movements with amplitudes $A=10^{\circ}$ and $40^{\circ}$, and the numerical results are shown in Fig. \ref{fig:New_Old_WSScomparisons}. The results show that the Newtonian model presents slightly lower average WSS values ($\lesssim 1.9\%$) in the maximum peak: for an amplitude $A=10^{\circ}$, the Newtonian model has a maximum peak of $7.23 \text{ Pa}$ while for the Oldroyd-B model with $\beta=0.95$ the value is $7.37 \text{ Pa}$; and for an amplitude $A=40^{\circ}$ the Newtonian model has a maximum peak of
$9.20 \text{ Pa}$ while for the Oldroyd-B model with $\beta=0.95$ the value is $9.38 \text{ Pa}$.
Similarly to the velocity magnitude results, there are no significant differences in the WSS results between the two different solvent viscosity ratios studied using the Oldroyd-B model.

Siluron 2000 is composed of two different MW molecules: 95\% of PDMS with a shear viscosity of $1 \text{ Pa s}$  and 5\% of PDMS with a shear viscosity of $2500 \text{ Pa s}$. The manufacturers argue that with the formulation used in Siluron 2000 its viscous properties can be modified depending on the permanent high shear forces, such as those generated in the eye due to its constant movement and increase the fluid resistance to emulsification \citep{Fluoron}. The results presented here suggest that for the typical movements that the eye is subjected to, the differences in the flow behaviour between the Newtonian model and the Oldroyd-B model tested are very small. In any case, for all the remaining results shown in this study, Siluron 2000 is simulated as a viscoelastic fluid with $\beta=0.95$.
%The low elasticity of Siluron 2000 is not sufficient to affect significantly its flow behaviour during typical saccadic movements in the range of amplitudes tested.

\subsection{Comparison of the flow behaviour of the various pharmacological fluids}

This subsection presents a comparative study of the flow behaviour of the pharmacological fluids under study. Velocity and WSS profiles are presented and discussed for all the fluids under study, when subjected to a saccadic movement with amplitude $A= 40^{\circ}$.

\subsubsection{Velocity field}
\label{vel_contours_40degrees}

\begin{figure}[!ht]
	\centering
	{\includegraphics[scale=0.35]{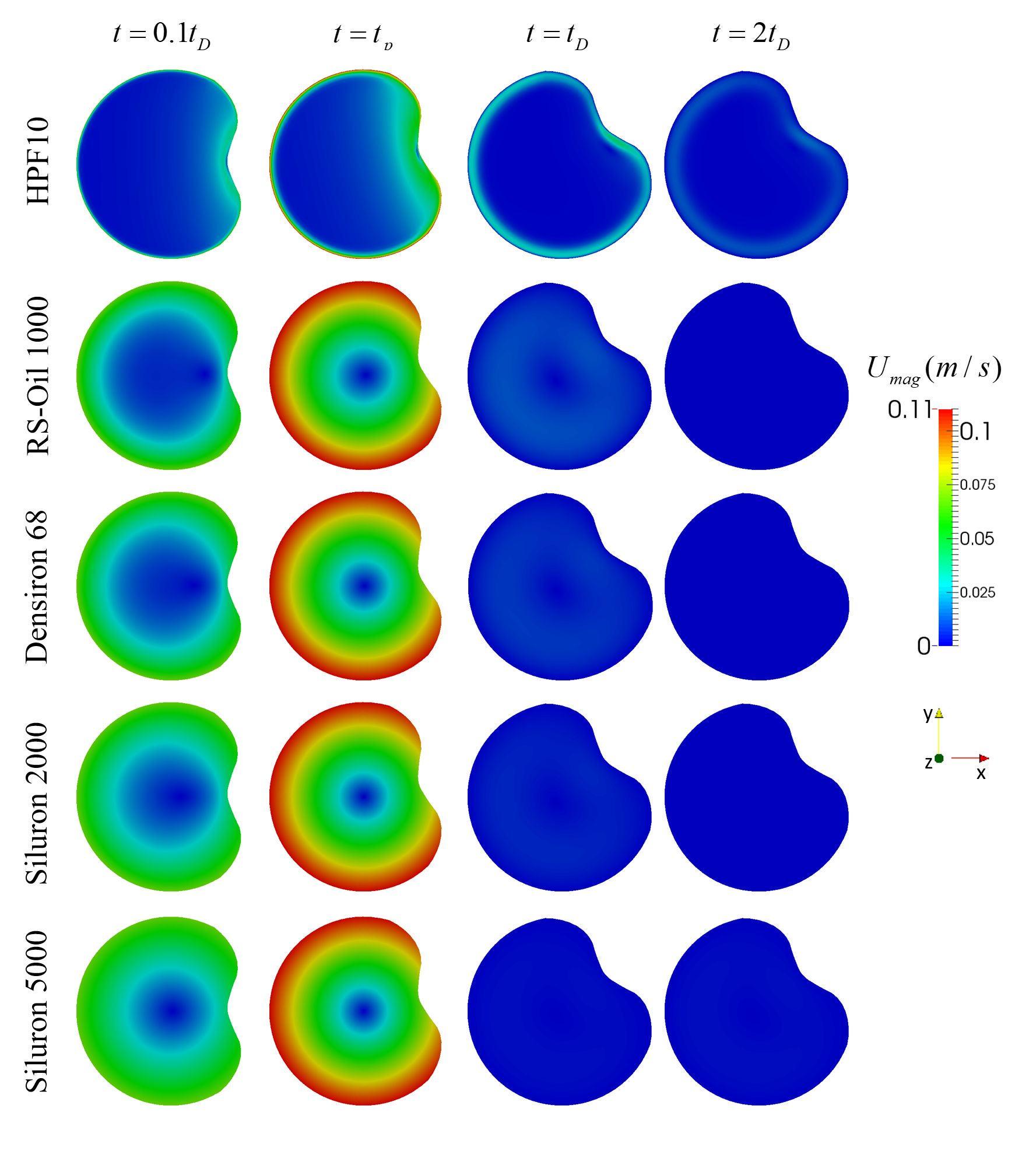}}
	\caption{Velocity magnitude contours on plane $z=0$ for a saccadic movement with amplitude $A=40^{\circ}$ at times $t=0.1t_D=0.0125 \text{ s}$, 
		$t=t_p=0.0375 \text{ s}$, $t=t_D=0.125 \text{ s}$ and $t=2t_D=0.25 \text{ s}$, for all pharmacological fluids under study.}
	\label{fig:vel_contours_all}
\end{figure}

The velocity magnitude contours in the plane $z=0$ for all the fluids under study are presented in Fig. \ref{fig:vel_contours_all}. The vitreous cavity rotates anti-clockwise around the $z$ axis. HPF10 presents very low viscosity (at $T=37 \text{ } ^{\circ}\text{C}$, the viscosity is $3.57\times10^{-3} \text{ Pa s}$), which means that the diffusive time scale is much higher than for the remaining fluids. Considering the eye as a perfect sphere with a diameter of $D=0.024 \text{ m}$, the diffusive time scale can be estimated as $t_{diff}=\frac{\rho D^2}{\eta}$: HPF10 shows a diffusive time scale of 303 s, while for all the SiO-based fluids the diffusive time scale is lower than 1 s. Note that the values are not accurate for the real geometry under study, but they still provide an estimate of the differences in the diffusive time scales between the tested fluids. 
%Comparing the two PFLC under study, the diffusive time scale of the fluid HPF10 is much lower than that of HPF8 and the momentum diffuses faster in the vitreous cavity due to the HPF10 viscosity being much higher than the HPF8 viscosity. 
For the HPF10 fluid, the lens indentation significantly affects the flow dynamics: for times $t=0.1t_D$ and $t=t_p$ the velocity magnitude right behind the lens indentation is considerable when compared to the rest of the  vitreous cavity. Finally, it is possible to observe that for $t=2t_D$, due to inertial effects, there is still fluid motion close to the walls.

The SiO-based fluids show distinct velocity profiles when compared to the PFLC. Due to their high shear viscosity, the diffusive time scale is low and for $t=t_p$ 
there is fluid motion across the whole vitreous cavity, except at the centre. It is possible to consider that these fluids move similarly to a rigid body, where individual fluid particles display a rotational motion, without being significantly deformed. With the increase of the SiO viscosity, the momentum propagates faster in the vitreous cavity (see the
velocity contours for $t=0.1t_D$). For time $t=t_D$ there is almost no fluid motion in the vitreous cavity and the flow stops right after $t=t_D$. In our previous work \citep{Silva2020} we show that for times higher than $t=t_D$ the VH gel phase produces considerable velocity gradients inside the vitreous cavity. Based on the results presented
in Fig. \ref{fig:vel_contours_all}, it seems that the lens indentation affects the velocity magnitude contours of the SiO-based fluids mostly in the beginning of the movement.

\cite{Abouali2012} also studied the flow behaviour of three different Newtonian fluids when subjected to saccadic movements with amplitudes between
$A=10 ^{\circ}$ and $50^{\circ}$. Their geometry was a simplified vitreous cavity and did not show the complexity of the vitreous cavity studied here. Nevertheless,
for a saccadic movement of $A=50^{\circ}$ the flow behaviour of water and a SiO with a shear viscosity of $0.97 \text{ Pa s}$ followed a similar trend to the results presented here.

\subsubsection{Wall shear stress}

\begin{figure}[!ht]
	\centering
	{\includegraphics[scale=0.25]{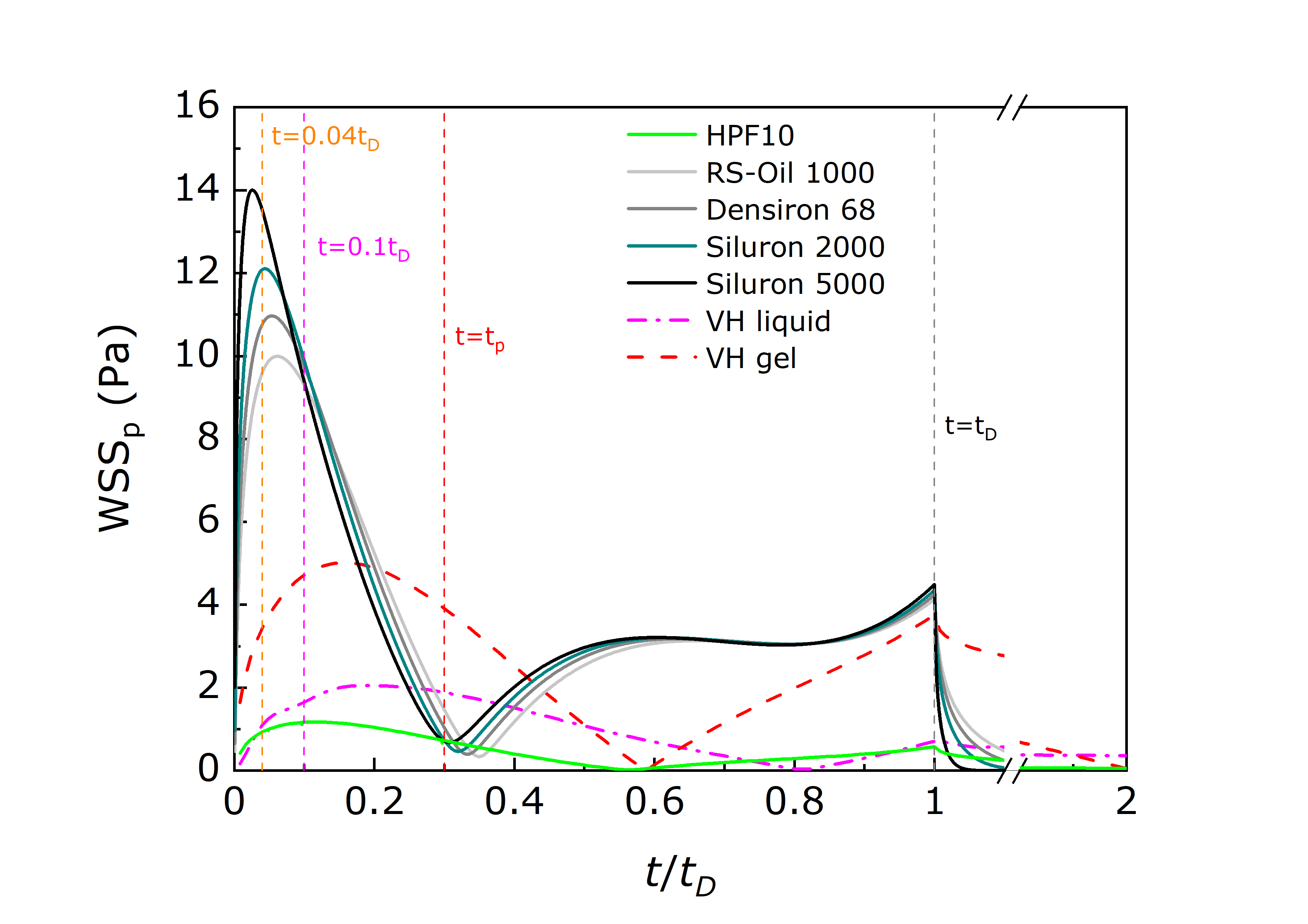}}
	\caption{Wall shear stress on point P for a saccadic amplitudes  $A=40^{\circ}$, for all the fluids under study. The WSS$_{\text{p}}$ for Vitreous Humour gel and liquids phase computed in our previous work \citep{Silva2020} are also presented.}
	\label{fig:WSS_over_time_all}
\end{figure}

The time variation of the wall shear stress at point P, WSS$_{\text{p}}$, for all the fluids under study is presented in Fig. \ref{fig:WSS_over_time_all}. Point P is located in the posterior position of  the  vitreous chamber and represents a position close to the center of the macula (point with  coordinates (-12,0,0) mm, see Supplementary Material 1). Note that for Siluron 2000 fluid the WSS include both the Newtonian and polymeric contributions. Additionally, the time and value of the maximum peak WSS$_{\text{p}}$ reached by each fluid is presented in Table \ref{tab:WSS_peak_values}. For the SiO-based fluids, the peak WSS$_{\text{p}}$ occurs at the beginning of the movement for $t<0.1t_D$, and these fluids present the highest values, between $10.0 \text{ Pa}$ for RS-Oil 1000 and $14.0 \text{ Pa}$ for Siluron 5000, according to their viscosity. 
All the SiOs present their lowest WSS$_{\text{p}}$ at $t=t_p$, or right after that time. The HPF10 presents a peak WSS$_{\text{p}}$ of $1.17 \text{ Pa}$ for $t/t_D=0.12$, and the lowest WSS$_{\text{p}}$ occurs at $t/t_D=0.56$.

\begin{table}[]
	\centering
	\caption{Maximum peak WSS at point P (WSS$_{\text{p}}$) during a saccadic movement with amplitude $A=40^{\circ}$.}
	\label{tab:WSS_peak_values}
	\begin{tabular}{ccc}
		\hline
		Fluid  & time (s)          & Peak WSS$_{\text{p}}$ (Pa)         \\ \hline
		HPF10                  & 0.120            & 1.17             \\
		RS-Oil 1000            & 0.0616            & 10.00            \\
		Densiron 68            & 0.0536            & 10.97             \\
		Siluron 2000           & 0.0440            & 12.11             \\
		Siluron 5000           & 0.0256            & 14.00            \\ \hline
	\end{tabular}
\end{table}

\begin{figure}[!ht]
	\centering
	{\includegraphics[scale=0.30]{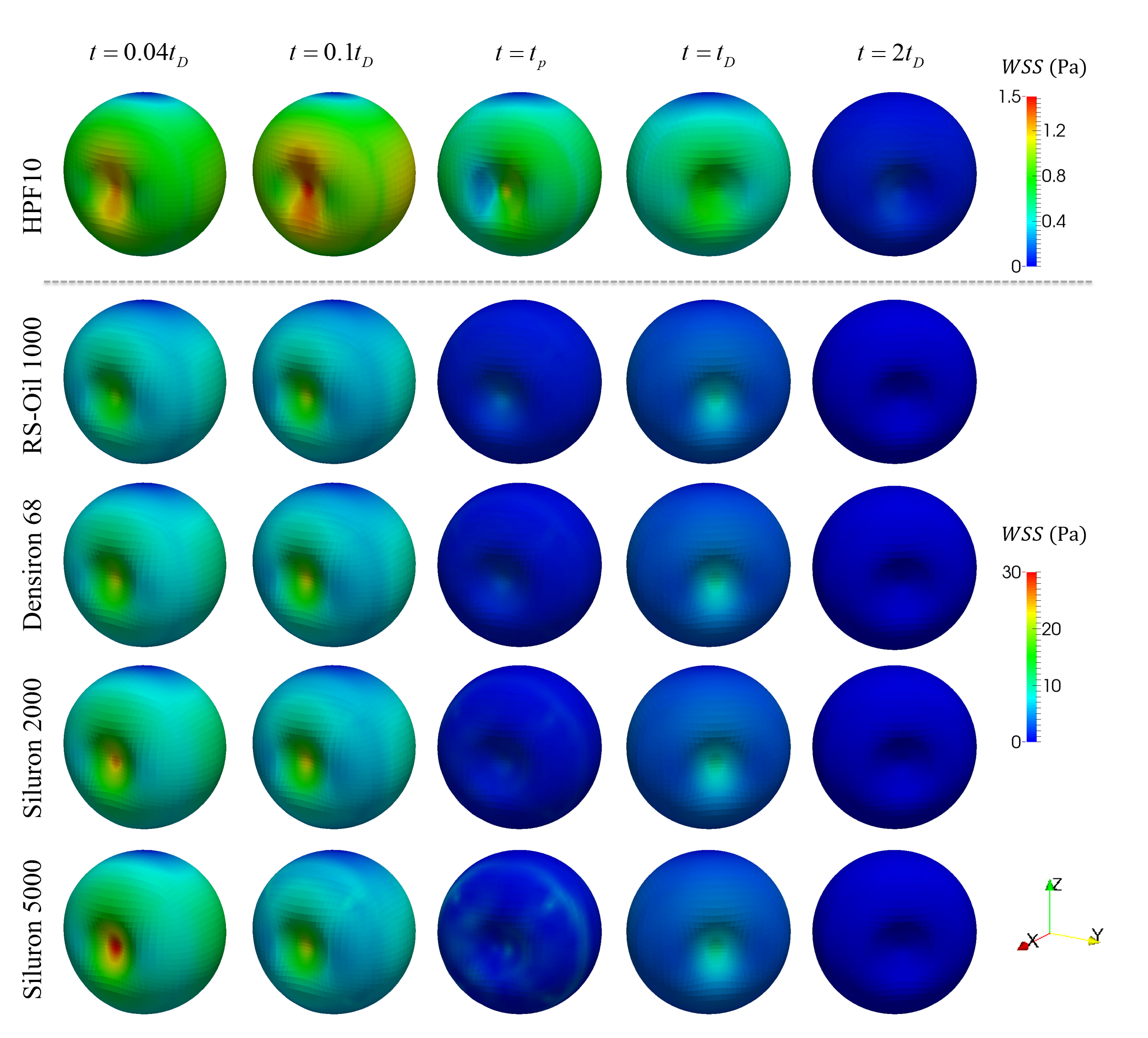}}
	\caption{Wall shear stress contours on the vitreous cavity for a saccadic movement with amplitude $A=40^{\circ}$ at times $t=0.04t_D=0.005 \text{ s}$, $t=0.1t_D=0.0125 \text{ s}$, 
		$t=t_p=0.0375 \text{ s}$, $t=t_D=0.125 \text{ s}$ and $t=2t_D=0.25 \text{ s}$, for all pharmacological fluids under study.}
	\label{fig:WSS_contours_all}
\end{figure}

The WSS contours of the pharmacological fluids under study for a saccadic movement of $A=40^{\circ}$ are presented in Fig. \ref{fig:WSS_contours_all}. Due to the
high WSS values reached in the beginning of the movement with the SiO-based fluids, the WSS contours for $t=0.04t_D$ are also presented. Note that due to the differences in the maximum WSS values, two different scales were used. The results show that, for all fluids tested, the maximum values of WSS occur around the centre of the lens indentation. 
The HPF10 fluid shows an interesting WSS profile along the lens indentation, with higher values concentrated along the $z$-axis reaching the outer area of the lens indentation
(see $t=0.04t_D$ and $t=0.1t_D$), whereas for all the other fluids the results show a constant WSS in the region around the lens indentation. For $t=t_p$, the fluid HPF10 also shows WSS contours lower in one side of the indentation ($y<0$) than in the other ($y>0$). The maximum WSS obtained for the HPF10 fluid is approximately $ 1.5 \text{ Pa}$. From all the fluids under study, SiO-based fluids show the highest WSS values. Siluron 5000 fluid reaches a maximum WSS in the centre of the lens indentation around $30 \text{ Pa}$, while RS-Oil 1000 reaches a maximum WSS around $18 \text{ Pa}$.

In the study performed by \cite{Abouali2012}, the WSS contours are presented for water, glycerol and a SiO fluid, and the values are in agreement with the ones presented here. However, in their study the WSS contours in the borders of the lens indentation show considerable differences when compared to the values inside and outside the lens 
indentation. Such phenomenon seems to be a consequence of the geometry in that area. Here, the geometry used presents a smoother transition between the lens indentation and the rest of the vitreous cavity, and no discrepancy was found between the WSS values on that region and in the neighbourhood.   
\par
The pharmacological fluids are mostly used in eye surgery to reattach the retina or close retinal tears that are created in the VH fluid as a consequence of the degradation of the collagen and HA structures. The forces generated between the fluid and the retina are key to keeping the retina in its proper position. The PFLC (HPF10) generates stresses much lower than the VH gel phase (see \cite{Silva2020}), which might compromise their effectiveness to reattach the retina or eliminate retinal tears. In fact, as mentioned in the introduction, the use of these fluids in eye surgery is rare, as SiO-based fluids are the main choice for the treatment of RD and RT. 
The SiO-based fluids generate stresses higher than VH gel phase, which seems appropriate to reattach the retina, as higher forces will pull and keep the retina in its proper position. However, it can be found in the literature that the use of SiO-based fluids for long periods often leads to complications such as cataracts, glaucoma, and corneal damage among others \citep{Baino2011,Giordano1998,Federman1988}. This can be related to the fact that the WSS values produced by the SiO are more than the double of those produced by VH gel phase (Siluron 2000 reaches a maximum WSS$_{\text{p}}$ of 12.11 Pa, while VH gel phase has a maximum WSS$_{\text{p}}$ of 5 Pa). From a mechanical point of view, a fluid with a behaviour similar to the VH gel phase, that would be able to produce stresses in the walls with similar magnitudes as the ones produced by VH gel phase, would be ideal to be used as a permanent VH substitute. 

\subsection{Impact of the amplitude of the saccadic movement on the fluid flow behaviour}
\label{sec:impact_diff_degrees}

A comparison of the velocity magnitude and WSS contours for different saccadic movement amplitudes, $A=10^{\circ}$, $20^{\circ}$, $30^{\circ}$ and $40^{\circ}$, are presented and discussed in this subsection. 
Results for HPF10 and Siluron 2000 are shown. Due to the similarities among the SiO-based fluids, the results RS-Oil 1000, Densiron 68 and Siluron 5000 fluids are not presented here for conciseness, but can be found in Supplementary Material 3.

\begin{figure}[!ht]
	\centering
	{\includegraphics[scale=0.3]{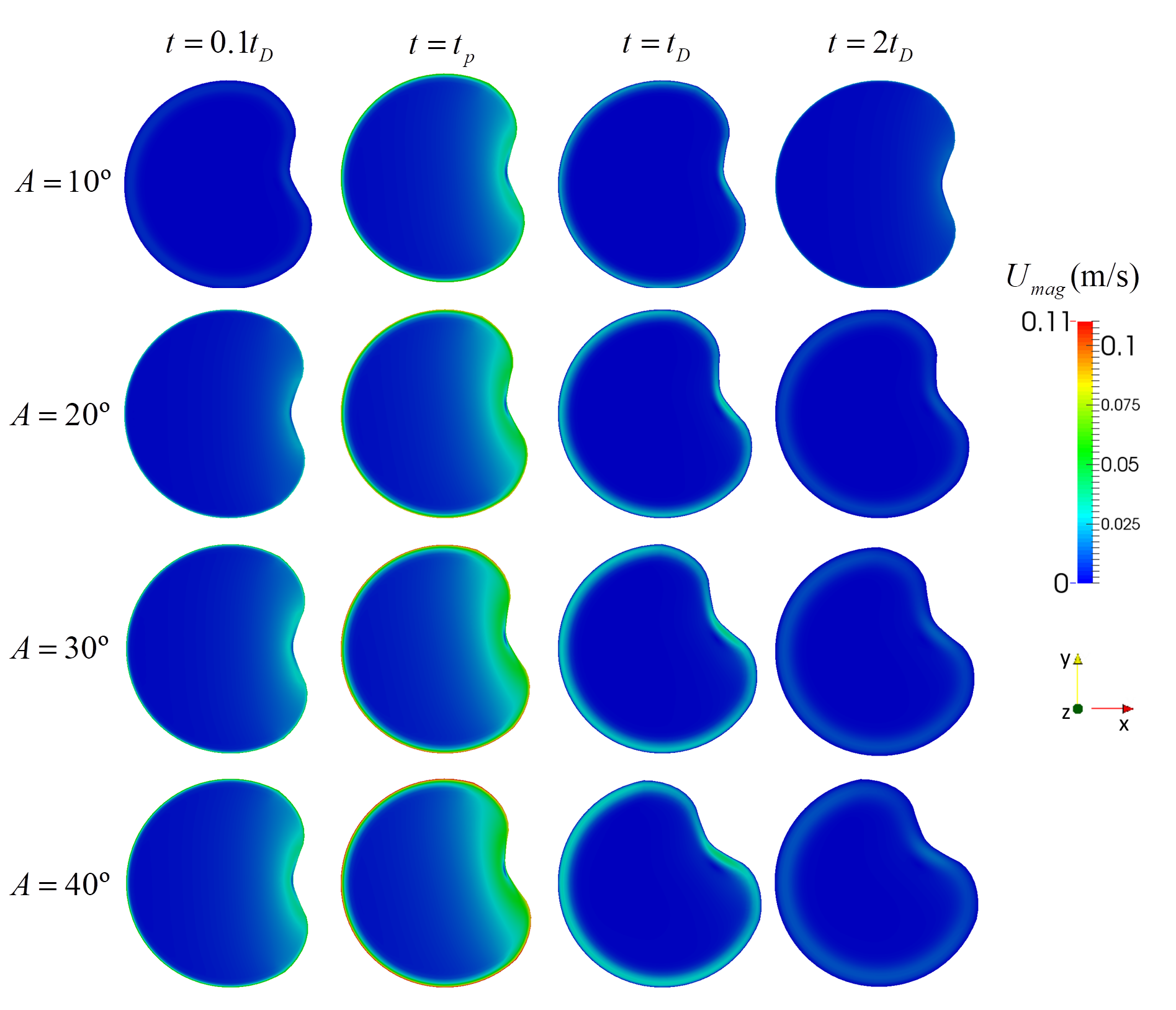}}
	\caption{Velocity magnitude contours on plane $z=0$ for saccadic movements with amplitudes $A=10^{\circ}$, $20^{\circ}$, $30^{\circ}$ and $40^{\circ}$, at times $t=0.1t_D$, $t=t_p$, $t=t_D$ and $t=2t_D$, for HPF10 fluid.}
	\label{fig:sac_amp_HPF10}
\end{figure}

\begin{figure}[!ht]
	\centering
	{\includegraphics[scale=0.3]{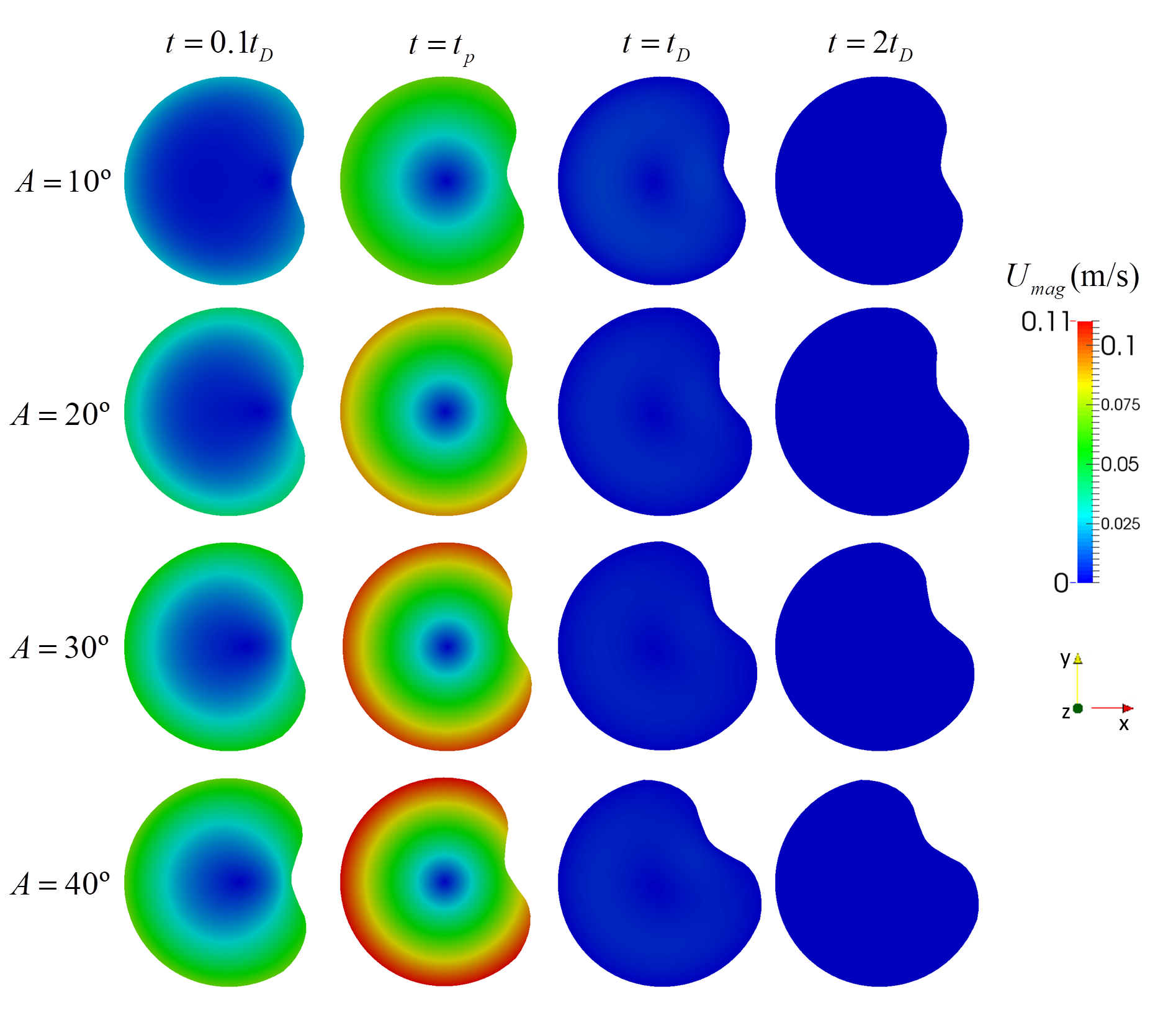}}
	\caption{Velocity magnitude contours on plane $z=0$ for saccadic movements with amplitudes $A=10^{\circ}$, $20^{\circ}$, $30^{\circ}$ and $40^{\circ}$, at times $t=0.1t_D$, $t=t_p$, $t=t_D$ and $t=2t_D$, for Siluron 2000 fluid.}
	\label{fig:sac_amp_Siluron2000}
\end{figure}

The velocity contours when the vitreous cavity is subjected to different amplitudes of movement, $A$, are presented in Fig. \ref{fig:sac_amp_HPF10} and Fig. \ref{fig:sac_amp_Siluron2000}, for HPF10 and Siluron 2000, respectively. The same trend is observed for both fluids when comparing different amplitudes of movement: with the increase of the saccadic amplitude, $A$, the angular velocity applied in the vitreous cavity walls also increases and, consequently, the velocities reached in the vitreous cavity increase.

\begin{figure}[!ht]
	\centering
	{\includegraphics[scale=0.25]{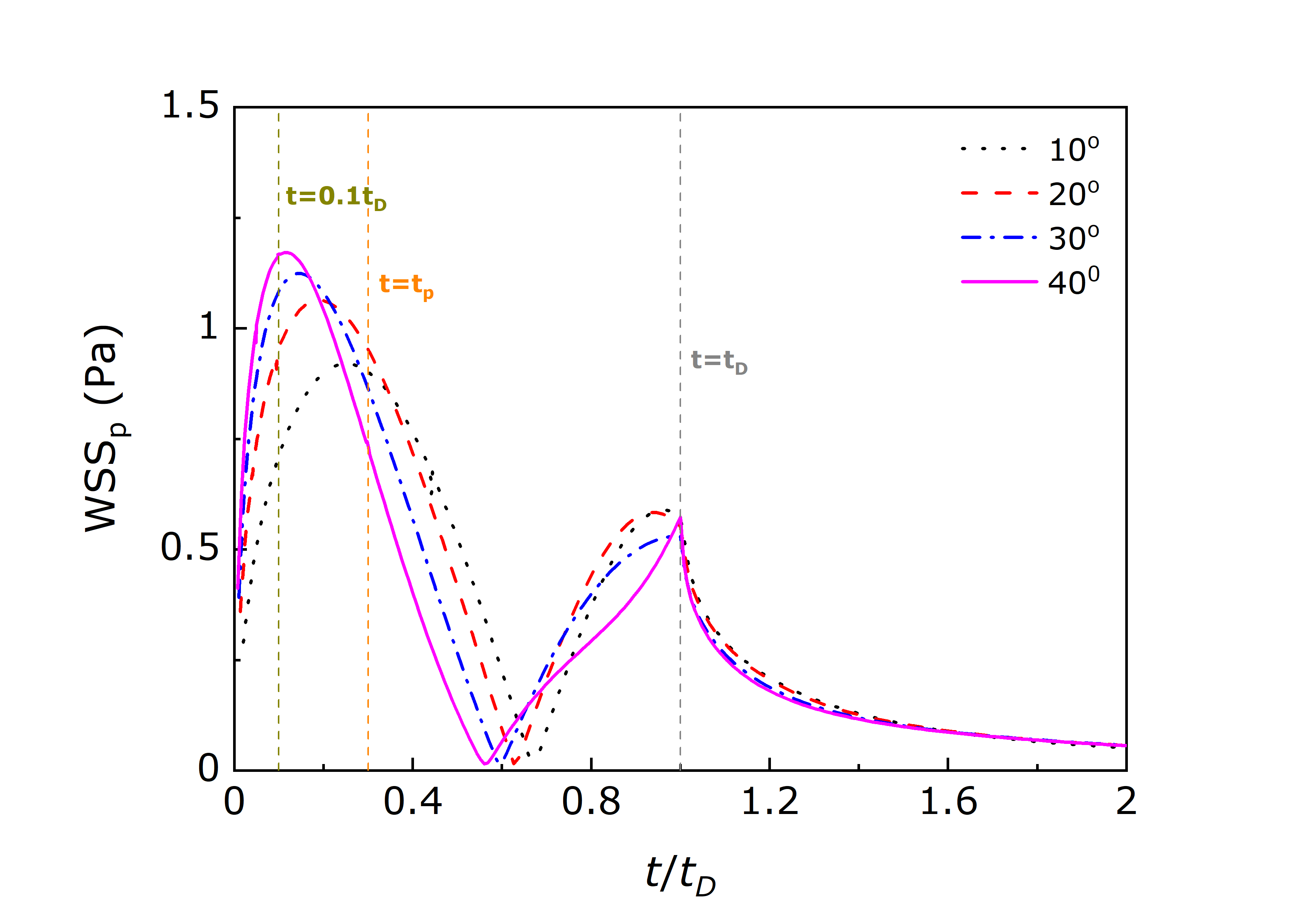}}
	\caption{Wall shear stress on point P for saccadic amplitudes of $A=10^{\circ}$, $20^{\circ}$, $30^{\circ}$ and $40^{\circ}$, for HPF10 fluid.}
	\label{fig:WSS_vs_time_diff_sac_deg_HPF10}
\end{figure}

\begin{figure}[!ht]
	\centering
	{\includegraphics[scale=0.315]{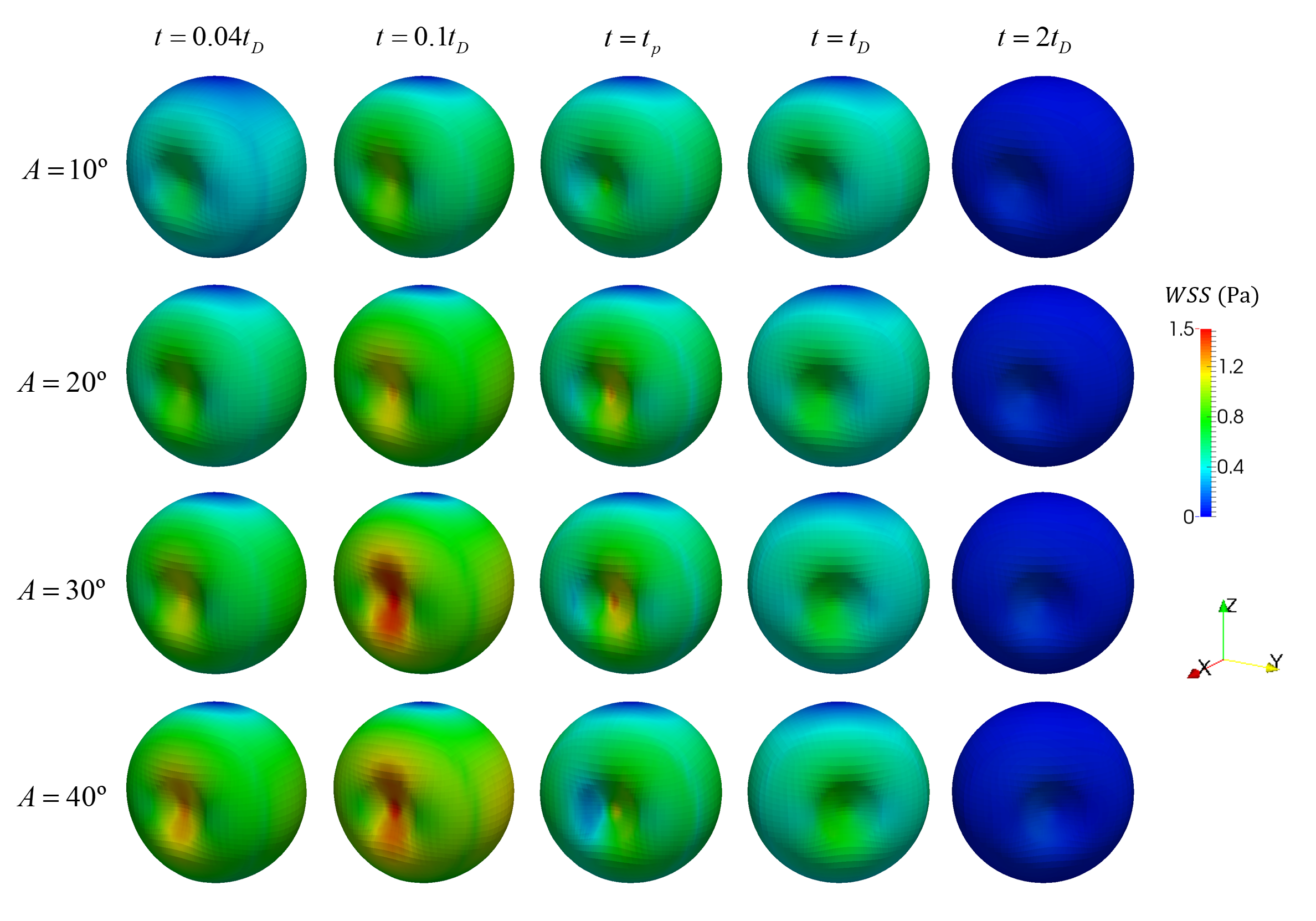}}
	\caption{Wall shear stress contours on the vitreous cavity for saccadic amplitudes of $A=10^{\circ}$, $20^{\circ}$, $30^{\circ}$ and $40^{\circ}$, at times $t=0.04t_D$, $t=0.1t_D$, $t=t_p$, $t=t_D$ and $t=2t_D$, for HPF10 fluid.}
	\label{fig:WSS_HPF10}
\end{figure}

The time variation of the WSS$_{\text{p}}$ in the vitreous cavity and the WSS contours for HPF10 fluid, for different saccade amplitudes and at different times, are shown in Fig.
\ref{fig:WSS_vs_time_diff_sac_deg_HPF10} and Fig. \ref{fig:WSS_HPF10}, respectively.
The numerical results show that the peak  WSS$_{\text{p}}$ increases with the increase of the saccadic amplitude up to a maximum of $1.17 \text{ Pa}$ for a saccadic movement with 
amplitude $A=40^{\circ}$. Also, with the increase of the saccadic movement amplitude, the peak value of WSS$_{\text{p}}$ is obtained for lower $t/t_D$. After reaching the peak value, 
for times between $t=0.1t_D$ and $t=t_p$, WSS$_{\text{p}}$ decreases to a minimum, and by the end of the saccadic movement it increases again reaching a similar value 
for all the amplitudes studied. For $t>t_D$ the WSS$_{\text{p}}$ follows the same trend and values for all the amplitudes under study. From the analysis
of Fig. \ref{fig:WSS_HPF10}, it is possible to observe that for $t=t_p$ the region of the lens indentation shows non-symmetric WSS values for all the degrees of movement under study, due to the important inertial effects in this low viscosity fluid.

\begin{figure}[!ht]
	\centering
	{\includegraphics[scale=0.25]{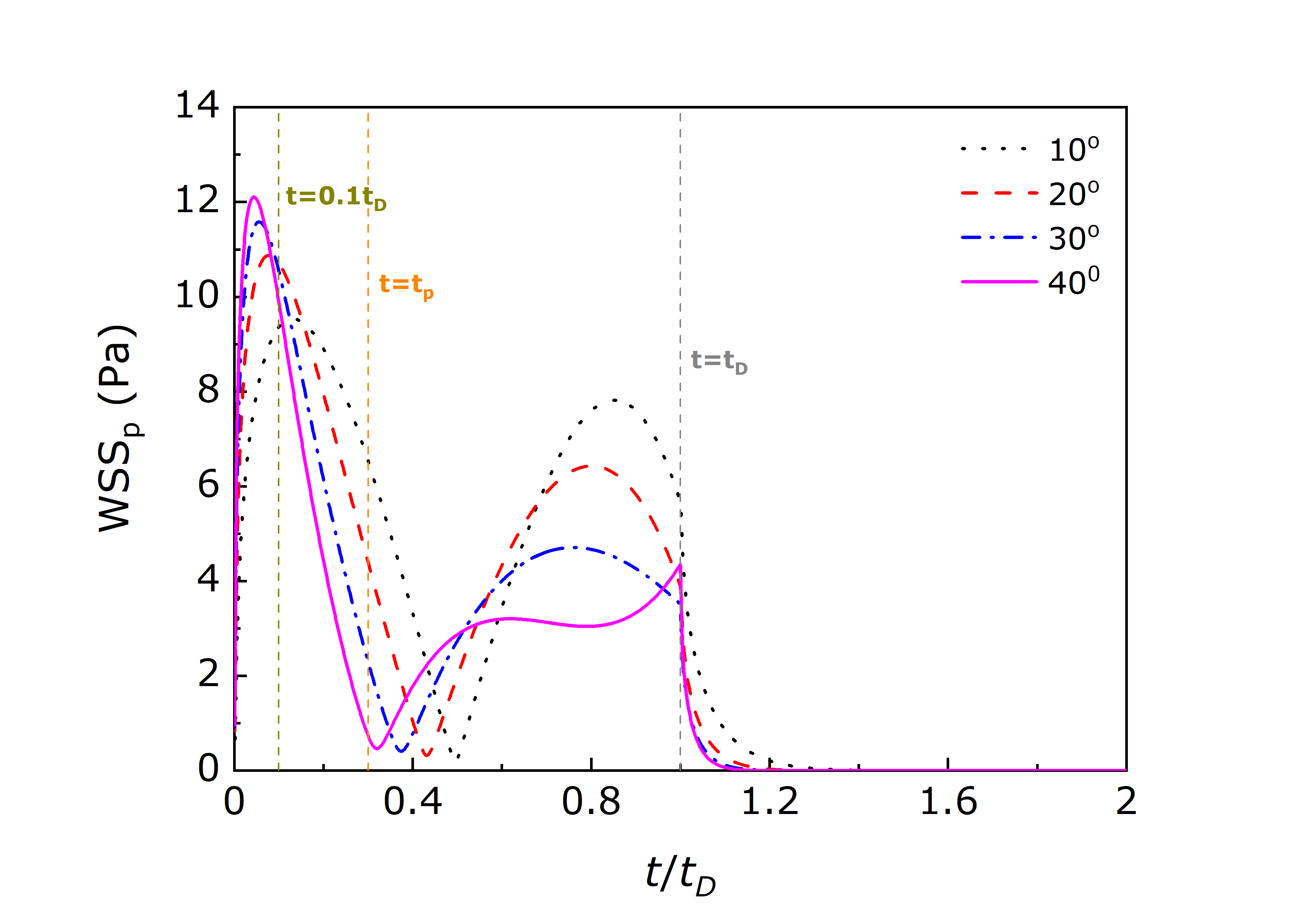}}
	\caption{Wall shear stress on point P for saccadic amplitudes of $A=10^{\circ}$, $20^{\circ}$, $30^{\circ}$ and $40^{\circ}$, for Siluron 2000 fluid.}
	\label{fig:WSS_vs_time_diff_sac_deg_Siluron2000}
\end{figure}

\begin{figure}[!ht]
	\centering
	{\includegraphics[scale=0.315]{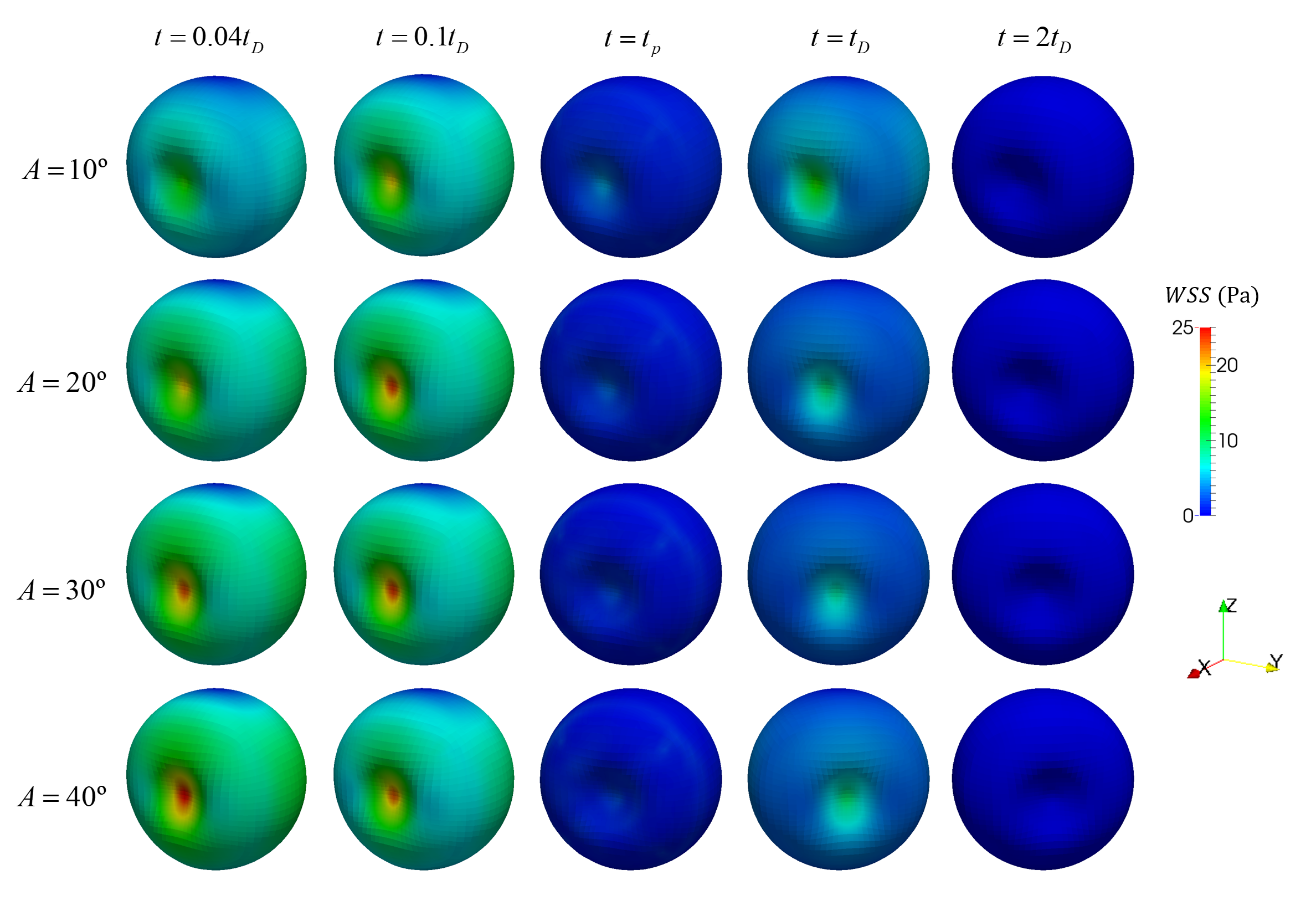}}
	\caption{Wall shear stress contours on the vitreous cavity for saccadic amplitudes of $A=10^{\circ}$, $20^{\circ}$, $30^{\circ}$ and $40^{\circ}$, at times $t=0.04t_D$, $t=0.1t_D$, $t=t_p$, $t=t_D$ and $t=2t_D$, for Siluron 2000 fluid.}
	\label{fig:WSS_Siluron2000}
\end{figure}

Figures \ref{fig:WSS_vs_time_diff_sac_deg_Siluron2000} and \ref{fig:WSS_Siluron2000} show respectively the time variation of  WSS$_{\text{p}}$ in the vitreous cavity and the WSS contours for Siluron 2000 fluid, for different amplitudes of movement and at different times. For $t<t_p$,  WSS$_{\text{p}}$ follows the same trend as observed for the HPF10 fluid, but the values are more than ten times higher. However, significant differences can be found for the second  WSS$_{\text{p}}$ peak: the movement with an amplitude of $A=10^{\circ}$ shows
a larger  WSS$_{\text{p}}$ peak than all the other movement amplitudes tested. Movements with amplitude $A=20$ and $30^{\circ}$  follow the same trend as the saccadic movement with 
$A=10^{\circ}$; however, a different profile is obtained for a movement with $A=40^{\circ}$. 
The difference in the WSS$_{\text{p}}$ profiles may be related to the saccadic profile applied (see the saccadic profiles in Supplementary material 2), as small saccades are characterised by a quasi-symmetric profile, while for higher saccadic amplitudes the deviations from a symmetric profile are considerable, and especially for a saccadic movement with $A=40^{\circ}$ the velocity profile in the final stages of the movement follows a different trend than all the other amplitudes tested. For all the amplitudes of movement studied, the maximum WSS is reached in the centre of the lens indentation and a smooth transition between the lens indentation area and the rest of the vitreous cavity is observed.

\section{Conclusions}

The main goal of this work was to characterise the rheological behaviour and quantify the differences in the flow behaviour of various vitreous substitutes available for eye surgery, and also the differences between those fluids and the physiological VH flow behaviour (see \cite{Silva2020}). 

PFLC and SiO fluids show distinct flow behaviour between them and also distinct flow behaviour when compared with VH \citep{Silva2020}. In fact, all the SiOs present WSS significantly higher than VH gel phase, while PFLC presents lower values. 

A comparison between the Siluron 2000 flow behaviour simulated as a low elasticity viscoelastic fluid and as a Newtonian fluid was also presented. The results showed that since
the elastic component of the Siluron 2000 fluid is weak, only minor differences were observed in the velocity and WSS profiles for eye movements tested. Therefore, it seems that during typical saccadic eye movements, the elastic component of Siluron 2000 fluid does not generate significant stresses due to elasticity and a behaviour similar to that of all the other SiO-based fluids that present a Newtonian rheology is observed.

None of the pharmacological fluids are able to mimic accurately the VH gel phase flow behaviour \citep{Silva2020}. We believe that a fluid with the aim of being used as a permanent VH substitute needs to properly mimic the VH gel flow behaviour besides having high stability and compatibility with the eye. The eye is constantly subjected to fast movements, among other movements, and finding a fluid that presents a similar behaviour with similar velocities and stresses as VH gel phase (the healthy VH conformation) is key to find a permanent VH substitute. 

\textbf{Acknowledgements:} A.F. Silva gratefully acknowledges the financial
support from the Funda\c{c}\~ao para a Ci\^encia e a Tecnologia (FCT) through scholarship SFRH/BD/91147/2012. M.S.N. Oliveira acknowledges the funding from Glasgow Research Partnership in Engineering (GRPe). The authors are grateful to Dr Patr\'{i}cia C. Sousa (Microfabrication and Exploratory Nanotechnology, INL) for advice and useful discussions about the extensional rheology results. Part of the results were obtained using the EPSRC funded ARCHIE-WeSt High Performance Computer (www.archiewest.ac.uk).

\bibliography{VH_substitutes}

\begin{thebibliography}{38}
\expandafter\ifx\csname natexlab\endcsname\relax\def\natexlab#1{#1}\fi
\providecommand{\url}[1]{\texttt{#1}}
\providecommand{\href}[2]{#2}
\providecommand{\path}[1]{#1}
\providecommand{\DOIprefix}{doi:}
\providecommand{\ArXivprefix}{arXiv:}
\providecommand{\URLprefix}{URL: }
\providecommand{\Pubmedprefix}{pmid:}
\providecommand{\doi}[1]{\href{http://dx.doi.org/#1}{\path{#1}}}
\providecommand{\Pubmed}[1]{\href{pmid:#1}{\path{#1}}}
\providecommand{\bibinfo}[2]{#2}
\ifx\xfnm\relax \def\xfnm[#1]{\unskip,\space#1}\fi
%Type = Article
\bibitem[{Abouali et~al.(2012)Abouali, Modareszadeh, Ghaffariyeh \&
  Tu}]{Abouali2012}
\bibinfo{author}{Abouali, O.}, \bibinfo{author}{Modareszadeh, A.},
  \bibinfo{author}{Ghaffariyeh, A.}, \& \bibinfo{author}{Tu, J.}
  (\bibinfo{year}{2012}).
\newblock \bibinfo{title}{Numerical simulation of the fluid dynamics in
  vitreous cavity due to saccadic eye movement}.
\newblock {\it \bibinfo{journal}{Medical Engineering and Physics}\/},  {\it
  \bibinfo{volume}{34}\/}, \bibinfo{pages}{681--92}.
  \DOIprefix\doi{10.1016/j.medengphy.2011.09.011}.
%Type = Article
\bibitem[{Alves et~al.(2003)Alves, Oliveira \& Pinho}]{Alves2003}
\bibinfo{author}{Alves, M.~A.}, \bibinfo{author}{Oliveira, P.~J.}, \&
  \bibinfo{author}{Pinho, F.~T.} (\bibinfo{year}{2003}).
\newblock \bibinfo{title}{A convergent and universally bounded interpolation
  scheme for the treatment of advection}.
\newblock {\it \bibinfo{journal}{International Journal for Numerical Methods in
  Fluids}\/},  {\it \bibinfo{volume}{41}\/}, \bibinfo{pages}{47--75}.
  \URLprefix \url{http://dx.doi.org/10.1002/fld.428}.
  \DOIprefix\doi{10.1002/fld.428}.
%Type = Article
\bibitem[{Baino(2010)}]{Baino2010}
\bibinfo{author}{Baino, F.} (\bibinfo{year}{2010}).
\newblock \bibinfo{title}{Scleral buckling biomaterials and implants for
  retinal detachment surgery}.
\newblock {\it \bibinfo{journal}{Medical Engineering \& Physics}\/},  {\it
  \bibinfo{volume}{32}\/}, \bibinfo{pages}{945--956}. \URLprefix
  \url{http://www.sciencedirect.com/science/article/pii/S1350453310001499}.
  \DOIprefix\doi{http://dx.doi.org/10.1016/j.medengphy.2010.07.007}.
%Type = Article
\bibitem[{Baino(2011)}]{Baino2011}
\bibinfo{author}{Baino, F.} (\bibinfo{year}{2011}).
\newblock \bibinfo{title}{Towards an ideal biomaterial for vitreous
  replacement: Historical overview and future trends}.
\newblock {\it \bibinfo{journal}{Acta Biomaterialia}\/},  {\it
  \bibinfo{volume}{7}\/}, \bibinfo{pages}{921--935}. \URLprefix
  \url{http://www.sciencedirect.com/science/article/pii/S1742706110005052}.
  \DOIprefix\doi{http://dx.doi.org/10.1016/j.actbio.2010.10.030}.
%Type = Article
\bibitem[{Barca et~al.(2014)Barca, Caporossi \& Rizzo}]{Barca2014}
\bibinfo{author}{Barca, F.}, \bibinfo{author}{Caporossi, T.}, \&
  \bibinfo{author}{Rizzo, S.} (\bibinfo{year}{2014}).
\newblock \bibinfo{title}{Silicone oil: Different physical proprieties and
  clinical applications}.
\newblock {\it \bibinfo{journal}{BioMed Research International}\/},  {\it
  \bibinfo{volume}{2014}\/}, \bibinfo{pages}{7}. \URLprefix
  \url{http://dx.doi.org/10.1155/2014/502143}.
  \DOIprefix\doi{10.1155/2014/502143}.
%Type = Book
\bibitem[{Becker(2006)}]{Becker1989}
\bibinfo{author}{Becker, W.} (\bibinfo{year}{2006}).
\newblock {\it \bibinfo{title}{The neurobiology of saccadic eye movements.
  Metrics}\/}.
\newblock \bibinfo{publisher}{PHI Learning}.
\newblock \URLprefix \url{https://books.google.co.uk/books?id=KEzdXmXgaHkC}.
%Type = Article
\bibitem[{Campo-Deano \& Clasen(2010)}]{Campo-Deano2010}
\bibinfo{author}{Campo-Deano, L.}, \& \bibinfo{author}{Clasen, C.}
  (\bibinfo{year}{2010}).
\newblock \bibinfo{title}{The slow retraction method (srm) for the
  determination of ultra-short relaxation times in capillary breakup
  extensional rheometry experiments}.
\newblock {\it \bibinfo{journal}{Journal of Non-Newtonian Fluid Mechanics}\/},
  {\it \bibinfo{volume}{165}\/}, \bibinfo{pages}{1688--1699}.
%Type = Article
\bibitem[{Caramoy et~al.(2011)Caramoy, Hagedorn, Fauser, Kugler, Gro{\ss} \&
  Kirchhof}]{Caramoy2011}
\bibinfo{author}{Caramoy, A.}, \bibinfo{author}{Hagedorn, N.},
  \bibinfo{author}{Fauser, S.}, \bibinfo{author}{Kugler, W.},
  \bibinfo{author}{Gro{\ss}, T.}, \& \bibinfo{author}{Kirchhof, B.}
  (\bibinfo{year}{2011}).
\newblock \bibinfo{title}{Development of emulsification-resistant silicone
  oils: Can we go beyond 2000 mpas silicone oil?}
\newblock {\it \bibinfo{journal}{Investigative Ophthalmology \& Visual
  Science}\/},  {\it \bibinfo{volume}{52}\/}. \URLprefix
  \url{http://dx.doi.org/10.1167/iovs.11-7250}.
  \DOIprefix\doi{10.1167/iovs.11-7250}.
%Type = Article
\bibitem[{Caramoy et~al.(2015)Caramoy, Kearns, Chan, Hagedorn, Poole, Wong,
  Fauser, Kugler, Kirchhof \& Williams}]{Caramoy2015}
\bibinfo{author}{Caramoy, A.}, \bibinfo{author}{Kearns, V.~R.},
  \bibinfo{author}{Chan, Y.~K.}, \bibinfo{author}{Hagedorn, N.},
  \bibinfo{author}{Poole, R.}, \bibinfo{author}{Wong, D.},
  \bibinfo{author}{Fauser, S.}, \bibinfo{author}{Kugler, W.},
  \bibinfo{author}{Kirchhof, B.}, \& \bibinfo{author}{Williams, R.}
  (\bibinfo{year}{2015}).
\newblock \bibinfo{title}{Development of emulsification resistant
  heavier-than-water tamponades using high molecular weight silicone oil
  polymers}.
\newblock {\it \bibinfo{journal}{Journal of Biomaterials Applications}\/},
  {\it \bibinfo{volume}{30}\/}, \bibinfo{pages}{212--220}. \URLprefix
  \url{https://doi.org/10.1177/0885328215575623}.
  \DOIprefix\doi{10.1177/0885328215575623}.
  \href{http://arxiv.org/abs/https://doi.org/10.1177/0885328215575623}{\tt
  arXiv:https://doi.org/10.1177/0885328215575623}.
\newblock \bibinfo{note}{PMID: 25766038}.
%Type = Article
\bibitem[{Caramoy et~al.(2010)Caramoy, Schröder, Fauser \&
  Kirchhof}]{Caramoy2010}
\bibinfo{author}{Caramoy, A.}, \bibinfo{author}{Schröder, S.},
  \bibinfo{author}{Fauser, S.}, \& \bibinfo{author}{Kirchhof, B.}
  (\bibinfo{year}{2010}).
\newblock \bibinfo{title}{In vitro emulsification assessment of new silicone
  oils}.
\newblock {\it \bibinfo{journal}{British Journal of Ophthalmology}\/},  {\it
  \bibinfo{volume}{94}\/}, \bibinfo{pages}{509--512}. \URLprefix
  \url{http://bjo.bmj.com/content/94/4/509.abstract}.
  \DOIprefix\doi{10.1136/bjo.2009.170852}.
%Type = Article
\bibitem[{Chan et~al.(2011)Chan, Ng, Knox, Garvey, Williams \& Wong}]{Chan2011}
\bibinfo{author}{Chan, Y.~K.}, \bibinfo{author}{Ng, C.~O.},
  \bibinfo{author}{Knox, P.~C.}, \bibinfo{author}{Garvey, M.~J.},
  \bibinfo{author}{Williams, R.~L.}, \& \bibinfo{author}{Wong, D.}
  (\bibinfo{year}{2011}).
\newblock \bibinfo{title}{Emulsification of silicone oil and eye movements}.
\newblock {\it \bibinfo{journal}{Investigative Ophthalmology \& Visual
  Science}\/},  {\it \bibinfo{volume}{52}\/}, \bibinfo{pages}{9721}. \URLprefix
  \url{+ http://dx.doi.org/10.1167/iovs.11-8586}.
  \DOIprefix\doi{10.1167/iovs.11-8586}.
  \href{http://arxiv.org/abs//data/journals/iovs/933457/z7g01311009721.pdf}{\tt
  arXiv:/data/journals/iovs/933457/z7g01311009721.pdf}.
%Type = Article
\bibitem[{Chirila et~al.(1998)Chirila, Hong, Dalton, Constable \&
  Refojo}]{Chirila1998}
\bibinfo{author}{Chirila, T.~V.}, \bibinfo{author}{Hong, Y.},
  \bibinfo{author}{Dalton, P.~D.}, \bibinfo{author}{Constable, I.~J.}, \&
  \bibinfo{author}{Refojo, M.~F.} (\bibinfo{year}{1998}).
\newblock \bibinfo{title}{The use of hydrophilic polymers as artificial
  vitreous}.
\newblock {\it \bibinfo{journal}{Progress in Polymer Science}\/},  {\it
  \bibinfo{volume}{23}\/}, \bibinfo{pages}{475--508}. \URLprefix
  \url{http://www.sciencedirect.com/science/article/pii/S0079670097000452}.
  \DOIprefix\doi{http://dx.doi.org/10.1016/S0079-6700(97)00045-2}.
%Type = Article
\bibitem[{Cibis et~al.(1962)Cibis, Becker, Okun \& Canaan}]{Cibis1962}
\bibinfo{author}{Cibis, P.~A.}, \bibinfo{author}{Becker, B.},
  \bibinfo{author}{Okun, E.}, \& \bibinfo{author}{Canaan, S.}
  (\bibinfo{year}{1962}).
\newblock \bibinfo{title}{The use of liquid silicone in retinal detachment
  surgery}.
\newblock {\it \bibinfo{journal}{Archives of Ophthalmology}\/},  {\it
  \bibinfo{volume}{68}\/}, \bibinfo{pages}{590--599}. \URLprefix \url{+
  http://dx.doi.org/10.1001/archopht.1962.00960030594005}.
  \DOIprefix\doi{10.1001/archopht.1962.00960030594005}.
%Type = Article
\bibitem[{Entov \& Hinch(1997)}]{Entov1997}
\bibinfo{author}{Entov, V.~M.}, \& \bibinfo{author}{Hinch, E.~J.}
  (\bibinfo{year}{1997}).
\newblock \bibinfo{title}{Effect of a spectrum of relaxation times on the
  capillary thinning of a filament of elastic liquid}.
\newblock {\it \bibinfo{journal}{Journal of Non-Newtonian Fluid Mechanics}\/},
  {\it \bibinfo{volume}{72}\/}, \bibinfo{pages}{31 -- 53}. \URLprefix
  \url{http://www.sciencedirect.com/science/article/pii/S0377025797000220}.
  \DOIprefix\doi{http://dx.doi.org/10.1016/S0377-0257(97)00022-0}.
%Type = Article
\bibitem[{Federman \& Schubert(1988)}]{Federman1988}
\bibinfo{author}{Federman, J.~L.}, \& \bibinfo{author}{Schubert, H.~D.}
  (\bibinfo{year}{1988}).
\newblock \bibinfo{title}{Complications associated with the use of silicone oil
  in 150 eyes after retina-vitreous surgery}.
\newblock {\it \bibinfo{journal}{Ophthalmology}\/},  {\it
  \bibinfo{volume}{95}\/}, \bibinfo{pages}{870 -- 876}. \URLprefix
  \url{http://www.sciencedirect.com/science/article/pii/S0161642088330800}.
  \DOIprefix\doi{http://dx.doi.org/10.1016/S0161-6420(88)33080-0}.
%Type = Article
\bibitem[{Feng et~al.(2013)Feng, Chen, Liu, Huang, Sun, Zhou, Lu \&
  Gao}]{Feng2013}
\bibinfo{author}{Feng, S.}, \bibinfo{author}{Chen, H.}, \bibinfo{author}{Liu,
  Y.}, \bibinfo{author}{Huang, Z.}, \bibinfo{author}{Sun, X.},
  \bibinfo{author}{Zhou, L.}, \bibinfo{author}{Lu, X.}, \&
  \bibinfo{author}{Gao, Q.} (\bibinfo{year}{2013}).
\newblock \bibinfo{title}{A novel vitreous substitute of using a foldable
  capsular vitreous body injected with polyvinylalcohol hydrogel}.
\newblock {\it \bibinfo{journal}{Scientific Reports}\/},  {\it
  \bibinfo{volume}{3}\/}. \URLprefix \url{<Go to ISI>://WOS:000318840800003}.
  \DOIprefix\doi{1838 10.1038/srep01838}.
%Type = Misc
\bibitem[{Fluoron(2016)}]{Fluoron}
\bibinfo{author}{Fluoron} (\bibinfo{year}{2016}).
\newblock \bibinfo{title}{Long-term tamponades, ultrapurified silicone oils and
  gases for intraocular use}.
\newblock \bibinfo{howpublished}{http://www.fluoron.de/}.
%Type = Article
\bibitem[{Giordano \& Refojo(1998)}]{Giordano1998}
\bibinfo{author}{Giordano, G.~G.}, \& \bibinfo{author}{Refojo, M.~F.}
  (\bibinfo{year}{1998}).
\newblock \bibinfo{title}{Silicone oils as vitreous substitutes}.
\newblock {\it \bibinfo{journal}{Progress in Polymer Science}\/},  {\it
  \bibinfo{volume}{23}\/}, \bibinfo{pages}{509--532}. \URLprefix
  \url{http://www.sciencedirect.com/science/article/pii/S0079670097000464}.
  \DOIprefix\doi{http://dx.doi.org/10.1016/S0079-6700(97)00046-4}.
%Type = Article
\bibitem[{Lee(1992)}]{Lee1992}
\bibinfo{author}{Lee, B.} (\bibinfo{year}{1992}).
\newblock \bibinfo{title}{Comparative rheological studies of the vitreous body
  of the eye}.
\newblock {\it \bibinfo{journal}{Dissertations available from ProQuest. Paper
  AAI9227704. http://repository.upenn.edu/dissertations/AAI9227704}\/}, .
%Type = Article
\bibitem[{Light(2006)}]{Light2006}
\bibinfo{author}{Light, D.~J.} (\bibinfo{year}{2006}).
\newblock \bibinfo{title}{Silicone oil emulsification in the anterior chamber
  after vitreoretinal surgery}.
\newblock {\it \bibinfo{journal}{Optometry - Journal of the American Optometric
  Association}\/},  {\it \bibinfo{volume}{77}\/}, \bibinfo{pages}{446 -- 449}.
  \URLprefix
  \url{//www.sciencedirect.com/science/article/pii/S1529183906003964}.
  \DOIprefix\doi{http://dx.doi.org/10.1016/j.optm.2006.04.119}.
%Type = Book
\bibitem[{Morrison(2001)}]{Morrison2001}
\bibinfo{author}{Morrison, F.~A.} (\bibinfo{year}{2001}).
\newblock {\it \bibinfo{title}{Understanding Rheology}\/}.
\newblock \bibinfo{publisher}{Oxford University Press}.
\newblock \URLprefix \url{https://books.google.co.uk/books?id=bwTn8ZbR0C4C}.
%Type = Article
\bibitem[{Nickerson et~al.(2005)Nickerson, Karageozian, Park \&
  Kornfield}]{Nickerson2005}
\bibinfo{author}{Nickerson, C.~S.}, \bibinfo{author}{Karageozian, H.~L.},
  \bibinfo{author}{Park, J.}, \& \bibinfo{author}{Kornfield, J.~A.}
  (\bibinfo{year}{2005}).
\newblock \bibinfo{title}{Internal tension: A novel hypothesis concerning the
  mechanical properties of the vitreous humor}.
\newblock {\it \bibinfo{journal}{Macromolecular Symposia}\/},  {\it
  \bibinfo{volume}{227}\/}, \bibinfo{pages}{183--190}. \URLprefix
  \url{http://dx.doi.org/10.1002/masy.200550918}.
  \DOIprefix\doi{10.1002/masy.200550918}.
%Type = Article
\bibitem[{Nickerson et~al.(2008)Nickerson, Park, Kornfield \&
  Karageozian}]{Nickerson2008}
\bibinfo{author}{Nickerson, C.~S.}, \bibinfo{author}{Park, J.},
  \bibinfo{author}{Kornfield, J.~A.}, \& \bibinfo{author}{Karageozian, H.}
  (\bibinfo{year}{2008}).
\newblock \bibinfo{title}{Rheological properties of the vitreous and the role
  of hyaluronic acid}.
\newblock {\it \bibinfo{journal}{Journal of Biomechanics}\/},  {\it
  \bibinfo{volume}{41}\/}, \bibinfo{pages}{1840--1846}. \URLprefix
  \url{http://www.sciencedirect.com/science/article/pii/S0021929008001917}.
  \DOIprefix\doi{http://dx.doi.org/10.1016/j.jbiomech.2008.04.015}.
%Type = Book
\bibitem[{Owens \& Phillips(2002)}]{Owens2002}
\bibinfo{author}{Owens, R.~G.}, \& \bibinfo{author}{Phillips, T.~N.}
  (\bibinfo{year}{2002}).
\newblock {\it \bibinfo{title}{Computational Rheology}\/}.
\newblock \bibinfo{publisher}{World Scientific Publishing Company}.
\newblock \URLprefix \url{https://books.google.co.uk/books?id=bPi3CgAAQBAJ}.
%Type = Article
\bibitem[{Papageorgiou(1995)}]{Papageorgiou1995}
\bibinfo{author}{Papageorgiou, D.~T.} (\bibinfo{year}{1995}).
\newblock \bibinfo{title}{On the breakup of viscous liquid threads}.
\newblock {\it \bibinfo{journal}{Physics of Fluids}\/},  {\it
  \bibinfo{volume}{7}\/}, \bibinfo{pages}{1529--1544}. \URLprefix
  \url{http://scitation.aip.org/content/aip/journal/pof2/7/7/10.1063/1.868540}.
  \DOIprefix\doi{doi:http://dx.doi.org/10.1063/1.868540}.
%Type = Article
\bibitem[{Pimenta \& Alves(2017)}]{Pimenta2017}
\bibinfo{author}{Pimenta, F.}, \& \bibinfo{author}{Alves, M.~A.}
  (\bibinfo{year}{2017}).
\newblock \bibinfo{title}{Stabilization of an open-source finite-volume solver
  for viscoelastic fluid flows}.
\newblock {\it \bibinfo{journal}{Journal of Non-Newtonian Fluid Mechanics}\/},
  {\it \bibinfo{volume}{239}\/}, \bibinfo{pages}{85 -- 104}. \URLprefix
  \url{http://www.sciencedirect.com/science/article/pii/S0377025716303329}.
  \DOIprefix\doi{http://dx.doi.org/10.1016/j.jnnfm.2016.12.002}.
%Type = Article
\bibitem[{Repetto et~al.(2005)Repetto, Stocchino \& Cafferata}]{Repetto2005}
\bibinfo{author}{Repetto, R.}, \bibinfo{author}{Stocchino, A.}, \&
  \bibinfo{author}{Cafferata, C.} (\bibinfo{year}{2005}).
\newblock \bibinfo{title}{Experimental investigation of vitreous humour motion
  within a human eye model}.
\newblock {\it \bibinfo{journal}{Physics in Medicine and Biology}\/},  {\it
  \bibinfo{volume}{50}\/}, \bibinfo{pages}{4729}. \URLprefix
  \url{http://stacks.iop.org/0031-9155/50/i=19/a=021}.
%Type = Article
\bibitem[{Romano et~al.(2009)Romano, Groenwald, Das, Stappler, Wong \&
  Heimann}]{Romano2009}
\bibinfo{author}{Romano, M.~R.}, \bibinfo{author}{Groenwald, C.},
  \bibinfo{author}{Das, R.}, \bibinfo{author}{Stappler, T.},
  \bibinfo{author}{Wong, D.}, \& \bibinfo{author}{Heimann, H.}
  (\bibinfo{year}{2009}).
\newblock \bibinfo{title}{Removal of densiron-68 with a 23-gauge
  transconjunctival vitrectomy system}.
\newblock {\it \bibinfo{journal}{Eye}\/},  {\it \bibinfo{volume}{23}\/},
  \bibinfo{pages}{715–717}.
  \DOIprefix\doi{http://dx.doi.org/10.1038/eye.2008.396}.
%Type = Article
\bibitem[{Sharif-Kashani et~al.(2011)Sharif-Kashani, Hubschman, Sassoon \&
  Kavehpour}]{Sharif-Kashani2011}
\bibinfo{author}{Sharif-Kashani, P.}, \bibinfo{author}{Hubschman, J.~P.},
  \bibinfo{author}{Sassoon, D.}, \& \bibinfo{author}{Kavehpour, H.~P.}
  (\bibinfo{year}{2011}).
\newblock \bibinfo{title}{Rheology of the vitreous gel: effects of
  macromolecule organization on the viscoelastic properties}.
\newblock {\it \bibinfo{journal}{Journal of Biomechanics}\/},  {\it
  \bibinfo{volume}{44}\/}, \bibinfo{pages}{419--23}.
  \DOIprefix\doi{10.1016/j.jbiomech.2010.10.002}.
%Type = Article
\bibitem[{Siggers \& Ethier(2012)}]{Siggers2012}
\bibinfo{author}{Siggers, J.~H.}, \& \bibinfo{author}{Ethier, C.~R.}
  (\bibinfo{year}{2012}).
\newblock \bibinfo{title}{Fluid mechanics of the eye}.
\newblock {\it \bibinfo{journal}{Annual Review of Fluid Mechanics}\/},  {\it
  \bibinfo{volume}{44}\/}, \bibinfo{pages}{347--372}. \URLprefix
  \url{http://www.annualreviews.org/doi/abs/10.1146/annurev-fluid-120710-101058}.
  \DOIprefix\doi{doi:10.1146/annurev-fluid-120710-101058}.
%Type = Article
\bibitem[{Silva et~al.(2017)Silva, Alves \& Oliveira}]{Silva2017}
\bibinfo{author}{Silva, A.~F.}, \bibinfo{author}{Alves, M.~A.}, \&
  \bibinfo{author}{Oliveira, M. S.~N.} (\bibinfo{year}{2017}).
\newblock \bibinfo{title}{Rheological behaviour of vitreous humour}.
\newblock {\it \bibinfo{journal}{Rheologica Acta}\/},  (pp.
  \bibinfo{pages}{1--10}). \URLprefix
  \url{http://dx.doi.org/10.1007/s00397-017-0997-0}.
  \DOIprefix\doi{10.1007/s00397-017-0997-0}.
%Type = Article
\bibitem[{Silva et~al.(2020)Silva, Pimenta, Alves \& Oliveira}]{Silva2020}
\bibinfo{author}{Silva, A.~F.}, \bibinfo{author}{Pimenta, F.},
  \bibinfo{author}{Alves, M.~A.}, \& \bibinfo{author}{Oliveira, M. S.~N.}
  (\bibinfo{year}{2020}).
\newblock \bibinfo{title}{Flow dynamics of vitreous humour during saccadic eye
  movements}.
\newblock {\it \bibinfo{journal}{Journal of the Mechanical Behavior of
  Biomedical Materials}\/},  {\it \bibinfo{volume}{110}\/},
  \bibinfo{pages}{103860}. \URLprefix
  \url{http://www.sciencedirect.com/science/article/pii/S1751616120304148}.
  \DOIprefix\doi{https://doi.org/10.1016/j.jmbbm.2020.103860}.
%Type = Article
\bibitem[{Sousa et~al.(2018)Sousa, Vaz, Cerejo, Oliveira, Alves \&
  Pinho}]{Sousa2018}
\bibinfo{author}{Sousa, P.~C.}, \bibinfo{author}{Vaz, R.},
  \bibinfo{author}{Cerejo, A.}, \bibinfo{author}{Oliveira, M. S.~N.},
  \bibinfo{author}{Alves, M.~A.}, \& \bibinfo{author}{Pinho, F.~T.}
  (\bibinfo{year}{2018}).
\newblock \bibinfo{title}{Rheological behavior of human blood in uniaxial
  extensional flow}.
\newblock {\it \bibinfo{journal}{Journal of Rheology}\/},  {\it
  \bibinfo{volume}{62}\/}, \bibinfo{pages}{447--456}. \URLprefix
  \url{https://doi.org/10.1122/1.4998704}. \DOIprefix\doi{10.1122/1.4998704}.
  \href{http://arxiv.org/abs/https://doi.org/10.1122/1.4998704}{\tt
  arXiv:https://doi.org/10.1122/1.4998704}.
%Type = Article
\bibitem[{Sousa et~al.(2017)Sousa, Vega, Sousa, Montanero \& Alves}]{Sousa2017}
\bibinfo{author}{Sousa, P.~C.}, \bibinfo{author}{Vega, E.~J.},
  \bibinfo{author}{Sousa, R.~G.}, \bibinfo{author}{Montanero, J.~M.}, \&
  \bibinfo{author}{Alves, M.~A.} (\bibinfo{year}{2017}).
\newblock \bibinfo{title}{Measurement of relaxation times in extensional flow
  of weakly viscoelastic polymer solutions}.
\newblock {\it \bibinfo{journal}{Rheologica Acta}\/},  {\it
  \bibinfo{volume}{56}\/}, \bibinfo{pages}{11--20}. \URLprefix
  \url{http://dx.doi.org/10.1007/s00397-016-0980-1}.
  \DOIprefix\doi{10.1007/s00397-016-0980-1}.
%Type = Article
\bibitem[{Stone(1958)}]{Stone1958}
\bibinfo{author}{Stone, W.~J.} (\bibinfo{year}{1958}).
\newblock \bibinfo{title}{Alloplasty in surgery of the eye}.
\newblock {\it \bibinfo{journal}{New England Journal of Medicine}\/},  {\it
  \bibinfo{volume}{258}\/}, \bibinfo{pages}{596--602}. \URLprefix
  \url{http://dx.doi.org/10.1056/NEJM195803202581206}.
  \DOIprefix\doi{10.1056/NEJM195803202581206}.
  \href{http://arxiv.org/abs/http://dx.doi.org/10.1056/NEJM195803202581206}{\tt
  arXiv:http://dx.doi.org/10.1056/NEJM195803202581206}.
\newblock \bibinfo{note}{PMID: 13517524}.
%Type = Article
\bibitem[{Vadillo et~al.(2012)Vadillo, Mathues \& Clasen}]{Vadillo2012}
\bibinfo{author}{Vadillo, D.~C.}, \bibinfo{author}{Mathues, W.}, \&
  \bibinfo{author}{Clasen, C.} (\bibinfo{year}{2012}).
\newblock \bibinfo{title}{Microsecond relaxation processes in shear and
  extensional flows of weakly elastic polymer solutions}.
\newblock {\it \bibinfo{journal}{Rheologica Acta}\/},  {\it
  \bibinfo{volume}{51}\/}, \bibinfo{pages}{755--769}. \URLprefix
  \url{http://dx.doi.org/10.1007/s00397-012-0640-z}.
  \DOIprefix\doi{10.1007/s00397-012-0640-z}.
%Type = Article
\bibitem[{Williams et~al.(2010)Williams, Day, Garvey, English \&
  Wong}]{Williams2010b}
\bibinfo{author}{Williams, R.~L.}, \bibinfo{author}{Day, M.},
  \bibinfo{author}{Garvey, M.~J.}, \bibinfo{author}{English, R.}, \&
  \bibinfo{author}{Wong, D.} (\bibinfo{year}{2010}).
\newblock \bibinfo{title}{Increasing the extensional viscosity of silicone oil
  reduces the tendency for emulsification}.
\newblock {\it \bibinfo{journal}{RETINA}\/},  {\it \bibinfo{volume}{30}\/},
  \bibinfo{pages}{--}. \URLprefix
  \url{http://journals.lww.com/retinajournal/Fulltext/2010/02000/INCREASING_THE_EXTENSIONAL_VISCOSITY_OF_SILICONE.15.aspx}.
%Type = Article
\bibitem[{Williams et~al.(2011)Williams, Day, Garvey, Morphis, Irigoyen, Wong
  \& Stappler}]{Williams2010}
\bibinfo{author}{Williams, R.~L.}, \bibinfo{author}{Day, M.~J.},
  \bibinfo{author}{Garvey, M.~J.}, \bibinfo{author}{Morphis, G.},
  \bibinfo{author}{Irigoyen, C.}, \bibinfo{author}{Wong, D.}, \&
  \bibinfo{author}{Stappler, T.} (\bibinfo{year}{2011}).
\newblock \bibinfo{title}{Injectability of silicone oil-based tamponade
  agents}.
\newblock {\it \bibinfo{journal}{British Journal of Ophthalmology}\/},  {\it
  \bibinfo{volume}{95}\/}, \bibinfo{pages}{273--276}. \URLprefix
  \url{http://bjo.bmj.com/content/95/2/273}.
  \DOIprefix\doi{10.1136/bjo.2010.192344}.
  \href{http://arxiv.org/abs/http://bjo.bmj.com/content/95/2/273.full.pdf}{\tt
  arXiv:http://bjo.bmj.com/content/95/2/273.full.pdf}.

\end{thebibliography}

\end{document}